\documentclass[lettersize,journal]{IEEEtran}
\usepackage{amsmath,amsfonts}
\usepackage{algorithmic}
\usepackage{algorithm}
\usepackage{array}

\usepackage{textcomp}
\usepackage{stfloats}
\usepackage{url}
\usepackage{verbatim}
\usepackage{graphicx}

\usepackage{amsthm,bm}
\usepackage{booktabs}
\usepackage{mathtools}
\usepackage{amssymb}
\usepackage{subcaption}
\usepackage{balance}

\newtheorem{remark}{Remark}
\usepackage[
backend=biber,
style=ieee,
sorting=none,
url=false,
doi=false,
isbn=false,
eprint=false
]{biblatex}
\usepackage{shellesc}

\newcounter{citebykeywordctr}
\newwrite\citebykeywordarg

\newif\ifcitebykeywordlive
\citebykeywordlivefalse

\newcommand{\citebykeyword}[1]{%
	\stepcounter{citebykeywordctr}%
	\edef\tempbase{tmpcite-\arabic{citebykeywordctr}}%
	\typeout{citebykeyword tempbase = \tempbase}%
	\ifcitebykeywordlive
		\immediate\openout\citebykeywordarg=\tempbase.arg\relax
		\immediate\write\citebykeywordarg{\detokenize{#1}}%
		\immediate\closeout\citebykeywordarg
		\ShellEscape{%
			python cite_by_keyword.py references.bib
			\tempbase.arg > \tempbase.tex
		}%
	\fi
	\IfFileExists{\tempbase.tex}
		{\input{\tempbase.tex}}
		{\textbf{Missing citation file: \tempbase.tex}}%
	\unskip\ignorespaces
}

\AtEveryBibitem{%
	\clearfield{note}%
}

\begin{document}

\title{Analytical Prediction of Equilibrium Existence in Current-Limited Grid-Forming Inverters}
\author{Wenhao Lin, Robin Preece, and Panagiotis N. Papadopoulos%
\thanks{All authors are with the Department of Electrical and Electronic Engineering, The University of Manchester, Manchester, UK (e-mail: wenhao.lin@postgrad.manchester.ac.uk; \{robin.preece, panagiotis.papadopoulos\}@manchester.ac.uk).}

\thanks{This work was supported by the UK Research and Innovation (UKRI) Future Leaders Fellowship under Grant MR/Y00390X/1 (P. N. Papadopoulos), and by the EPSRC-funded Supergen Energy Networks Hub under Grant EP/S00078X/2 (R. Preece).}

\thanks{All results can be fully reproduced using the methods and data described in this paper and the cited references. For the purpose of open access, the authors have applied a Creative Commons Attribution (CC BY) licence to any Author Accepted Manuscript version arising from this submission.}}

\markboth{}%
{Analytical Prediction of Equilibrium Existence in Current-Limited Grid-Forming Inverters}


\maketitle

\begin{abstract}
As converter-interfaced units become dominant in power systems, understanding nonlinear stability phenomena involving discrete switching events becomes essential for secure operation. The limited overcurrent capability of grid-forming (GFM) inverters makes current limiting essential during large disturbances. Activation of a circular current limiter (CCL) can cause the operating equilibrium to disappear under certain circumstances. This paper develops an analytical framework to predict the grid-voltage boundaries for CCL activation and determine whether the operating equilibrium persists or disappears after activation. For CCL-based GFM inverters with frozen anti-windup, closed-form expressions for the activation boundaries are derived while accounting for the filter capacitor. A criterion is established to determine whether a saturated (i.e., current-limited) stable equilibrium point (satSEP) persists following limiter activation. A continuation method is also developed to trace equilibria across normal-control and current-limited operation. Power-angle analysis and dynamic-model continuation in single-inverter and modified 9-bus systems validate these predictions. Time-domain simulations further confirm the results. The results show that CCL activation can either cause immediate loss of the stable equilibrium or allow stable current-limited operation to continue until the equilibrium is lost as operating conditions change further.
\end{abstract}


\section{Introduction}
Climate-change mitigation and net-zero targets are accelerating the replacement of fossil-fuel-based generation by renewable energy resources. These resources are predominantly interfaced with the grid through power-electronic converters, commonly referred to as inverter-based resources (IBRs). Most existing IBRs operate in grid-following mode. This operating mode is effective in strong grids, but it becomes less suitable as power systems become increasingly converter dominated \cite{nerc_integrating_2017}. Grid-forming (GFM) inverters are therefore expected to provide voltage and frequency support in power systems \cite{nerc_white_2023}.

Unlike synchronous generators (SGs), which can tolerate large short-term overcurrents \cite{de_metz-noblat_calculation_2005}, power-electronic converters have limited overcurrent capability and therefore require current-limiting control during severe voltage disturbances \cite{baeckeland_overcurrent_2024}. Common approaches for GFM converters include virtual impedance, which modifies the converter voltage--current relationship \citebykeyword{VICurrentLimiting}, and current-reference saturation (CRS), which limits the reference produced by the voltage-control loop before it enters the inner current controller.

Among CRS schemes, the widely used circular current limiter (CCL) limits the current magnitude while preserving the angle of the unsaturated reference \citebykeyword{CCL}. Alternatives include constant-angle CRS (CACRS), which imposes a prescribed angle \citebykeyword{FixAngle}, and priority-based CRS, which prioritizes either the \(d\)- or \(q\)-axis current \citebykeyword{priority-based current limiter}. It remains unclear which CRS scheme offers superior overall performance. Given the widespread use of CCL in GFM control studies, this paper focuses on CCL-based GFM inverters, while stability analyses of other CRS schemes can be found in \citebykeyword{dPriorityStability,CACRSStability}.

For CCL-based GFMs, an equivalent-circuit model and power-angle analysis were used in \cite{fan_equivalent_2022} to determine whether a stable equilibrium point (SEP) remains in the unsaturated region following grid-voltage variations. An SEP that exists while the current limiter is active is hereafter referred to as a saturated SEP (satSEP). Such satSEPs have been identified and analysed for CACRS-based current limiting \citebykeyword{CACRSStability}. However, the formation and loss of satSEPs in CCL-based GFMs remain insufficiently understood. Existing equivalent-circuit analyses also often neglect the filter-capacitor branch. Although neglecting the filter capacitor simplifies the power-angle expression, it may reduce the accuracy of the predicted current-limiting boundary, particularly during grid overvoltage, when the capacitor’s reactive-power injection increases the inverter’s reactive-power absorption requirement and causes the current limit to be reached earlier than predicted.

Hybrid and piecewise-smooth (PWS) dynamic analysis methods have been used to analyse switching events and controller limits in power systems, including transformer tap changes, generator PV–PQ switching due to reactive power limits, and excitation limits in automatic voltage regulators \cite{hiskens_trajectory_2000,geng_second-order_2019,reddy_limit-induced_2005}. These studies highlight the importance of accounting for changes in operating modes when assessing power-system stability. More recently, these approaches have been extended to inverter-based systems, with studies examining repeated current saturation, PWS bifurcations, and oscillatory behaviours in current-limited inverters \citebykeyword{PWS}. However, existing PWS studies rarely consider the CCL and have not analytically characterised its switching behaviour. A PWS formulation accounts for the change in inverter control when the CCL is activated, helping explain whether the inverter can continue operating under current limiting or loses its operating equilibrium.

To address these gaps, this paper investigates how satSEPs arise and disappear in a GFM inverter equipped with a CCL and frozen anti-windup. The inverter is formulated as a PWS system comprising the normal-control mode (NCM) and current-limited mode (CLM), and CCL activation is studied within a boundary equilibrium bifurcation (BEB) framework which includes two cases: persistence, where the equilibrium continues into the CLM, and a non-smooth saddle-node bifurcation, where the equilibrium disappears at the switching boundary. An equivalent-circuit model that includes the filter capacitor is developed to analytically predict CCL activation and the existence and loss of satSEPs. Using the derived analytical expressions, we determine whether the operating equilibrium persists in the current-limited mode or disappears upon CCL activation.

The main contributions of this paper are summarised as follows.
\begin{enumerate}
    \item This paper derives new analytical expressions for the lower and upper grid voltages at which the CCL is activated. These closed-form expressions are obtained from an extended equivalent-circuit model that includes the filter capacitor, improving prediction accuracy by accounting for its current contribution. They allow the onset of current limiting to be predicted directly from the active-power setpoint, voltage reference, current limit, and circuit parameters.

    \item This paper establishes an analytical criterion to predict whether the inverter loses its operating equilibrium when the CCL is activated or can continue operating in CLM. The criterion is derived from the slope of the saturated power-angle curve and can be evaluated directly from the system parameters, enabling rapid assessment without repeated time-domain simulations.    
    
    \item An analytical expression is derived for the collapse voltage at which the inverter loses its operating equilibrium. The slope condition identifies collapse at CCL activation, while the maximum transferable power determines when a persisting satSEP disappears. The resulting expression allows the collapse voltage to be calculated directly from the system parameters.
    
    \item A numerical continuation method is developed to validate the analytical predictions more efficiently by avoiding repeated time-domain simulations. The method switches from the normal-control model to the reduced current-limited model at CCL activation and continues tracing the equilibrium branch. It directly locates limiter activation and subsequent equilibrium loss with reduced computational effort.
\end{enumerate}

The rest of this paper is organised as follows. Section~\ref{sec:beb_scope} defines the boundary-equilibrium bifurcations. Section~\ref{sec:PWS_CCL} formulates the CCL-based GFM inverter as a PWS system and presents the continuation method that switches to a reduced CLM model at CCL activation. Section~\ref{sec:Analytical_Conditions} develops the equivalent-circuit model and derives the current-limiting boundaries, satSEP existence condition, and collapse voltage. Section~\ref{sec:application} applies these results to numerical examples and validates the predictions using power-angle analysis, dynamic-model continuation, and time-domain simulations. Section~\ref{sec:conclusion} concludes the paper.

\section{Boundary-Equilibrium Bifurcations}
\label{sec:beb_scope}

\begin{figure}[t]
	\centering
	\includegraphics[width=0.9\linewidth]{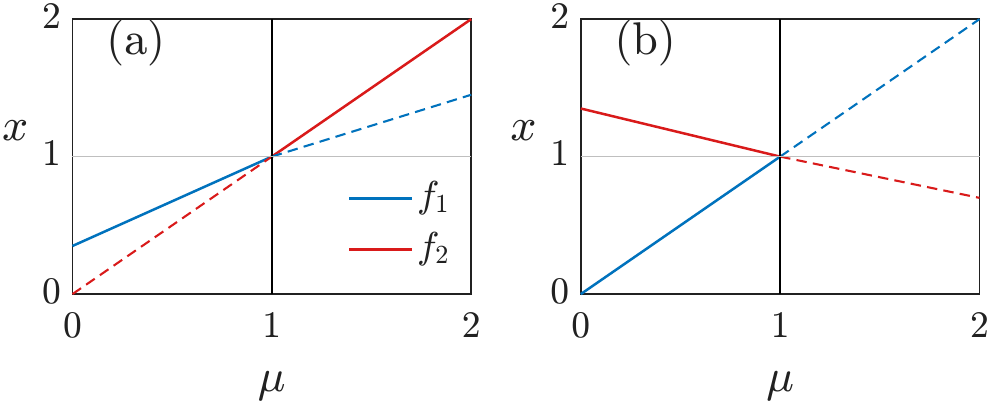}
	\caption{Canonical BEB scenarios at \(\mu=1\). Blue and red denote the equilibrium branch of the two subsystem equations; solid segments are admissible and dashed segments are virtual: (a) persistence and (b) a non-smooth fold. The line style indicates admissibility, not stability.}
	\label{fig:bebdemogroupcolors}
\end{figure}

Current limiting partitions the converter model into two smooth subsystems separated by a switching boundary. As a parameter varies, an equilibrium of one subsystem may reach this boundary and undergo a BEB \cite{di_bernardo_piecewise-smooth_2008}. A simple converter example is an increase in load, which raises the current reference magnitude towards the current limit. In this context, \(x\) can be interpreted as the magnitude of the current reference before limiting, and \(\mu\) as a load parameter. Figure~\ref{fig:bebdemogroupcolors} illustrates the two BEB arrangements, with the curves representing equilibrium solutions as \(\mu\) varies. The switching boundary is \(x=1\), corresponding to the current limit. In this scenario, a load of \(\mu=1\) results in a current reference magnitude of \(x=1\), placing the BEB at \((x,\mu)=(1,1)\).

Here, \(f_1\) and \(f_2\) can be interpreted as the converter dynamics without and with current limiting, respectively. The blue and red curves represent their equilibrium solutions. Each subsystem is active only in its own region. Therefore, an equilibrium is admissible if it lies in the region where the corresponding subsystem is active; otherwise, it is virtual. For example, in Fig.~\ref{fig:bebdemogroupcolors}(a), the subsystem without current limiting governs \(x<1\), so a solution of \(f_1=0\) is admissible for \(x<1\) but virtual for \(x>1\). 

Figure~\ref{fig:bebdemogroupcolors}(a) shows persistence BEB. As \(\mu\) passes through the BEB (i.e., $\mu>1$), the admissible equilibrium continues from the blue curve to the red curve. Therefore, the equilibrium does not disappear at the switching boundary. For the GFM inverter, the NCM equilibrium reaches the CCL boundary and continues into the CLM, forming a satSEP. The inverter will operate stably. Figure~\ref{fig:bebdemogroupcolors}(b) shows a non-smooth fold. After the non-smooth fold BEB, both equilibrium curves lie outside their corresponding active regions and therefore become virtual. Thus, no admissible equilibrium remains. For the GFM inverter, this means that the operating SEP is lost when the CCL is activated, and no satSEP is formed. The inverter will lose stability.

\section{Piecewise-Smooth Modelling of Current-Limited Inverter Systems}\label{sec:PWS_CCL}
This section formulates the current-limited inverter as a PWS system and proposes an event-triggered bifurcation-analysis framework for efficient satSEP validation.
\subsection{GFM Control Structure}
\label{sec:gfm_control_structure}
\begin{figure}[t]
    \centering
    \includegraphics[width=1\linewidth]{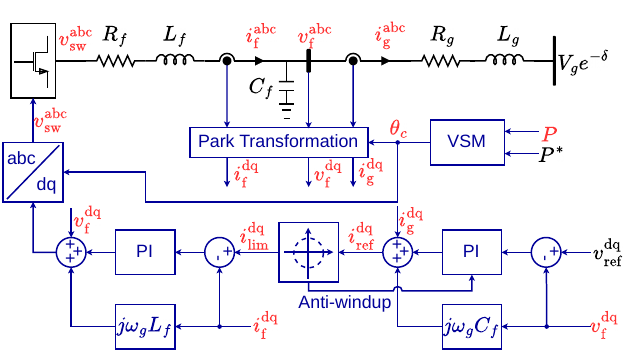}
    \caption{Control structure of the studied GFM inverter with circular current limiting and frozen anti-windup.}
    \label{fig:gfm_control_structure}
\end{figure}

The GFM inverter considered in this paper employs virtual synchronous machine (VSM) and cascaded outer voltage and inner current-control loops, following the control structure in \cite{fan_equivalent_2022}. Figure~\ref{fig:gfm_control_structure} shows the control structure and its connection to the inverter and filter.The subscripts $f$ and $g$ denote filter-related and grid-related quantities, respectively, while $v_{\rm sw}$ denotes the equivalent voltage of the inverter switching stage.

The voltage controller generates the unconstrained current reference \(i_{\rm ref}\), which passes through the CCL before being supplied to the inner current controller as \(i_{\rm lim}\). A frozen anti-windup scheme holds the voltage-controller integral states constant during current-limited operation.

\subsection{Operating Modes and CCL Activation}
\label{sec:circular_current_limiting}
The CCL constrains the magnitude of the current reference generated by the voltage controller while preserving its direction. When the reference magnitude exceeds the prescribed current limit, the CCL scales the reference vector to this limit. The inverter operates in the NCM when the CCL is inactive and in the CLM when it is active. Let
\begin{equation}
	i^{\rm dq}_{\rm ref} =
	\begin{bmatrix}
		i_{\rm ref}^{\rm d} & i_{\rm ref}^{\rm q}
	\end{bmatrix}^{\mathsf T}
\end{equation}
denote the unconstrained current reference. The switching function $h(x,p)$ and switching surface $\Sigma$ are
\begin{equation}
	h(x,p)=\|i^{\rm dq}_{\rm ref}(x,p)\|_2-\overline{I},
	\quad
	\Sigma=\{(x,p):h(x,p)=0\}.
	\label{eq:switching_surface}
\end{equation}
\(\overline{I}\) denotes the maximum continuous operating current, while $\boldsymbol{x}$ and $\boldsymbol{p}$ denote the system states and parameters, respectively. The NCM applies for \(h\leq0\), whereas the CLM applies for \(h>0\). The CCL and frozen anti-windup law are
\begin{equation}
	i^{\rm dq}_{\rm lim}
	=
	\begin{cases}
		i^{\rm dq}_{\rm ref},
		& h(x,p)\leq 0, \\[2mm]
		\overline{I}\dfrac{i^{\rm dq}_{\rm ref}}{|i^{\rm dq}_{\rm ref}|},
		& h(x,p)>0,
	\end{cases}
	\label{eq:circular_limiter}
\end{equation}
and
\begin{equation}
	\dot{\xi}^{v}
	=
	\begin{cases}
		V_{\mathrm{ref}}-V_f,
		& h(x,p)\leq 0,
		\\[1mm]
		0,
		& h(x,p)>0,
	\end{cases}
	\label{eq:frozen_integrator}
\end{equation}
respectively. Here, $\xi^{v}$ denotes the integral state of the voltage controller. The complete switched model is therefore
\begin{equation}
	\dot{x}
	=
	F_{\rm PWS}(x,p)
	=
	\begin{cases}
		F_{\rm NCM}(x,p),
		& h(x,p)\leq 0,
		\\[1mm]
		F_{\rm CLM}(x,p),
		& h(x,p)>0.
	\end{cases}
	\label{eq:piecewise_limited_model}
\end{equation}
Limiter activation changes the active vector field but does not reset the dynamic states. Thus, at the switching time \(t_s\),
\begin{equation}
	x(t_s^+)=x(t_s^-).
	\label{eq:state_continuity_switching}
\end{equation}
The same continuity holds for the equilibrium transition at the limiting event. Let \((x_s,p_s)\) be an NCM equilibrium on \(\Sigma\), such that
\begin{equation}
	F_{\rm NCM}(x_s,p_s)=0,
	\qquad
	h(x_s,p_s)=0.
	\label{eq:unsat_switching_equilibrium}
\end{equation}
On \(\Sigma\), the CCL reduces to the identity. Moreover, the NCM equilibrium satisfies \(V_{\mathrm{ref}}-V_f=0\), so the two expressions in \eqref{eq:frozen_integrator} also coincide at \((x_s,p_s)\). Consequently,
\begin{equation}
	F_{\rm CLM}(x_s,p_s)=0.
	\label{eq:sat_equilibrium_same_point}
\end{equation}
The NCM and CLM equilibrium branches therefore meet at the same event point. The voltages and currents of converter remain continuous when the current limiter is activated. These values provide the starting point for tracing operation in the current-limited mode.

\subsection{Reduced CLM Formulation for Equilibrium Continuation}
\label{sec:reduced_saturated_formulation}

The PWS model in \eqref{eq:piecewise_limited_model} is retained for time-domain simulation because it automatically selects the appropriate control mode at each time step. Although the two equilibrium branches meet at the event point, the Jacobian of the equilibrium equations is generally discontinuous there. Therefore, pseudo-arclength continuation applied directly to the complete PWS model may fail to pass through the switching event and follow the intended branch.

The two modes are consequently continued as separate segments. The NCM branch is followed until the event \(h=0\) is located. Continuation is then stopped, and the same event equilibrium is used to start the CLM segment. In this way, each continuation run uses one fixed set of equations, while \(h=0\) determines when the equations must be changed.

A second difficulty arises in the \(F_{\rm CLM}(x,p)\) equilibrium formulation. In the CLM, frozen anti-windup imposes
\begin{equation}
	\dot{\xi}_{v,d}=0,
	\qquad
	\dot{\xi}_{v,q}=0.
	\label{eq:zero_integrator_dynamics}
\end{equation}
If \(\xi_{v,d}\) and \(\xi_{v,q}\) are retained as states, the equilibrium Jacobian contains two zero rows and becomes rank deficient, preventing regular Newton correction and continuation.

The limiting event supplies physically consistent values for these frozen states. They are fixed at
\begin{equation}
	\xi_{v,d}=\bar{\xi}_{v,d},
	\qquad
	\xi_{v,q}=\bar{\xi}_{v,q},
	\label{eq:frozen_integrator_parameters}
\end{equation}
where the barred quantities are inherited from the NCM equilibrium at limiter activation. Removing the two frozen states and their zero equations gives the reduced CLM equilibrium problem
\begin{equation}
	F_{\rm CLM,red}
	\left(x_{\rm red};p,\bar{\xi}_{v,d},\bar{\xi}_{v,q}\right)=0,
	\label{eq:reduced_saturated_system}
\end{equation}
where \(x_{\rm red}\in\mathbb{R}^{n-2}\) for the original \(n\)-state model. The reduction removes only the frozen integrator states; all remaining converter and network dynamics are retained.

\section{Analytical Conditions for Equilibria in the Current-Limited Mode}\label{sec:Analytical_Conditions}
\begin{figure*}[!t]
	\normalsize
	\begin{equation}
		\bar{R}_e
		=
		\frac{
			\sqrt{
				A^2 V_{\mathrm{ref}}^2
				+ A V_g^2
				- H^2 \overline{I}^2
				+ 2 V_{\mathrm{ref}} V_g A
				\left[
				\left(B_c X_g - 1\right)\cos\delta
				+ B_c R_g \sin\delta
				\right]
			}
			-
			R_g \overline{I}
		}{
			\overline{I} A
		}.
		\label{eq:Re_bar}
	\end{equation}
	
	\begin{equation}
	P(\delta,V_g)
	=
	\frac{
		\left[
		\left(X_g B_c - 1\right)R_e - R_g
		\right]
		V_{\mathrm{ref}}V_g\cos\delta
		+
		V_{\mathrm{ref}}V_g
		\left(
		R_e B_c R_g + X_g
		\right)
		\sin\delta
		+
		V_{\mathrm{ref}}^2
		\left(
		A R_e + R_g
		\right)
	}{
		A R_e^2 + 2R_eR_g + Z
	}
	-
	R_e \overline{I}^2 .
	\label{eq:P_with_cap}
\end{equation}
	
	\begin{equation}
	R_e=\max\{0,\operatorname{Re}(\bar{R}_e)\},\quad Z = R_g^2 + X_g^2,
	\quad
	A = B_c^2 Z - 2B_c X_g + 1,
	\quad
	H = X_g - B_c Z . \label{eq:define_AHZ}
\end{equation}
	\vspace{-1em}
\end{figure*}

\begin{figure}[t]
	\centering
	\includegraphics[width=\linewidth]{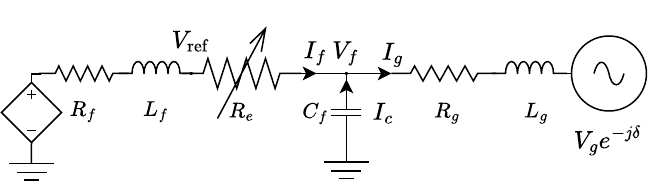}
	\caption{Equivalent-circuit representation of a current-limited inverter.}
	\label{fig:equivalendcircuit}
\end{figure}
This section uses an equivalent circuit to analyse the resulting CLM branch. The formulation gives the CLM power-angle characteristic, the boundary voltages at which the CCL is activated, the condition for satSEP existence, and the collapse voltage.

\subsection{Equivalent Circuit Model with Capacitor Effects}

The equivalent-circuit model formulation in \cite{fan_equivalent_2022} represents the action of the current limiter through a variable effective resistance \(R_e\), which modifies the inverter voltage--current relationship once the current limit is reached. However, the model neglects the current flowing through the filter capacitor. The filter capacitor supplies reactive power locally, reducing the converter current required during reactive-power injection but increasing it during reactive-power absorption, thereby delaying and advancing CCL activation, respectively. Accounting for this effect improves the accuracy of the predicted boundary voltages.

    Equations~\eqref{eq:Re_bar}--\eqref{eq:P_with_cap} are derived from circuit analysis of the equivalent circuit in Fig.~\ref{fig:equivalendcircuit}, accounting for the filter-capacitor current and current limiter. They express the effective virtual resistance \(R_e\) and the corresponding active-power as functions of \(\delta\) and \(V_g\). Here, \(\delta\) is the angle of the inverter voltage reference relative to the equivalent grid voltage, and \(V_{\mathrm{ref}}\) and \(V_g\) are their respective magnitudes. The grid reactance is \(X_g=\omega_g L_g\), where \(\omega_g\) is the grid angular frequency. The filter-capacitor susceptance is \(B_c=\omega_g C_f\), where \(C_f\) is the filter capacitance. The parameter \(\overline{I}\) denotes the maximum allowable inverter current. The parameters \(R_g\) and \(L_g\) denote the equivalent grid resistance and inductance, respectively. The quantities \(A\), \(H\), and \(Z\) are auxiliary coefficients introduced for notational convenience and defined in \eqref{eq:define_AHZ}.

Setting \(B_c=0\) gives \(A=1\) and \(H=X_g\), so \eqref{eq:Re_bar}--\eqref{eq:P_with_cap} reduce to the formulation in \cite{fan_equivalent_2022}, which neglects the filter capacitor.

\begin{remark}
	\label{rem:saturated_steady_state_approximation}
	For the analytical equivalent-circuit model, the contribution of the frozen voltage-controller integrator is approximated as negligible:
	\begin{equation}
		\xi_v\approx 0,
		\qquad
		\dot{\xi}_v=0.
		\label{eq:xi_steady_state_zero}
	\end{equation}
	This approximation is motivated by the feedforward compensation in the voltage-control loop, which reduces the steady-state contribution required from the integrator. It is used only in the analytical equivalent-circuit formulation; the reduced continuation model in Section~\ref{sec:reduced_saturated_formulation} retains the integrator values inherited at limiter activation. The accuracy of the approximation is subsequently assessed through bifurcation analysis of the dynamic model.
\end{remark}

\begin{remark}
	In the single-inverter infinite-bus (SIIB) system of Fig.~\ref{fig:equivalendcircuit}, \(V_g\) is the infinite-bus voltage. In a larger network, \(V_g\) denotes the Thevenin equivalent voltage seen from the inverter terminal rather than the voltage of a particular physical bus. The following analysis therefore applies to both the SIIB system and a network represented by its local Thevenin equivalent, as demonstrated for the multi-bus systems in Section~\ref{sec:result}.
\end{remark}

The equivalent-circuit characteristic is piecewise defined through \(R_e\). From the definition \(R_e=\max\{0,\operatorname{Re}(\bar{R}_e)\}\), the value of \(R_e\) depends directly on the sign of \(\operatorname{Re}(\bar{R}_e)\). When \(\operatorname{Re}(\bar{R}_e)>0\), a positive effective resistance is required to enforce the current limit, corresponding to the CLM. When \(\operatorname{Re}(\bar{R}_e)\leq 0\), no additional effective resistance is required and \(R_e\) is therefore set to zero, corresponding to the NCM. Hence,
\begin{equation}
	P(\delta,V_g)
	=
	\begin{cases}
		P_{\mathrm{NCM}}(\delta,V_g),
		& \operatorname{Re}\!\left(\bar{R}_e\right)\leq 0,\\[2mm]
		P_{\mathrm{CLM}}(\delta,V_g),
		& \operatorname{Re}\!\left(\bar{R}_e\right)>0,
	\end{cases}
	\label{eq:P_piecewise_cap}
\end{equation}
where \(P_{\mathrm{NCM}}\) is obtained by setting \(R_e=0\) in \eqref{eq:P_with_cap}, while \(P_{\mathrm{CLM}}\) is obtained by setting \(R_e=\operatorname{Re}\!\left(\bar{R}_e\right)\).

\subsection{Boundary Voltage}
The boundary-voltage analysis considers steady-state equilibria of the equivalent-circuit model with fixed circuit and control parameters. For a specified \(P^*\), the equilibrium angle \(\delta\) is determined by \(P(\delta,V_g)=P^*\) as \(V_g\) varies. At the limiting boundary, denoted by \((\delta_b,V_{g,b})\), the current magnitude equals \(\overline{I}\) and \(R_e=0\), giving
\begin{equation}
	P(\delta_b,V_{g,b})=P^*,
	\qquad
	\bar{R}_e=0,
	\label{eq:boundary_con}
\end{equation}

Solving \eqref{eq:boundary_con} gives
\begin{equation}
	V_{g,b}^{\pm}=\sqrt{
	A V_{\mathrm{ref}}^2
	+
	\overline{I}^2 Z
	-
	2P^*R_g
	\pm
	2H\sqrt{
		\overline{I}^2V_{\mathrm{ref}}^2-(P^*)^2
	}
}.
	\label{eq:Vgb}
\end{equation}
For \(H=X_g - B_c Z>0\), \(V_{g,b}^{-}\) and \(V_{g,b}^{+}\) are denoted as the lower boundary voltage (LBV) and upper boundary voltage (UBV), respectively. The LBV is encountered during a voltage decrease, when the GFM supplies increasing reactive current for voltage support, whereas the UBV is associated with reactive-power absorption. The inverter operates in NCM between the two admissible boundaries and enters CLM outside this range. Therefore, \(V_{g,b}^{\pm}\) provides an explicit threshold for predicting CCL activation during an increase or decrease in grid voltage.
\subsection{Condition for the Existence of satSEPs}
The boundary voltage identifies when the inverter reaches its current limit. If it cannot continue operating stably under current limiting, this is also the collapse voltage. To assess whether stable operation can continue, we evaluate the slope of the current-limited power-angle curve at this boundary:
\begin{equation}
	C(V_{g,b})=\left.
	\frac{\partial P_{\mathrm{CLM}}(\delta,V_g)}{\partial \delta}
	\right|_{\delta=\delta_b,V_g=V_{g,b}}
	>0 .
	\label{eq:SEP_condition}
\end{equation}
A positive slope means that the operating point meets the slope condition for a satSEP on the current-limited \(P\)-\(\delta\) curve, allowing stable operation to continue after limiter activation.
At the UBV, the boundary slope is
\begin{equation}
	C^+=C(V_{g,b}^+)=
	-\frac{
		P^*\!\left(
		A\sqrt{\overline{I}^{2}V_{\mathrm{ref}}^{2}-(P^*)^{2}}
		+H \overline{I}^{2}
		\right)
	}{
		\overline{I}^{2}R_g
	}.\label{eq:cond_Vgb_plus}
\end{equation}
Since \(A>0\), \(H>0\), \(R_g>0\), and \(P^*>0\), \(C^+\) is always negative. Hence, no satSEP exists beyond the UBV, and the operating equilibrium is lost at this boundary.

At the LBV, the boundary slope is
\begin{equation}
	C^-=C(V_{g,b}^-)=
	\frac{
		P^*
		\left(
		A \sqrt{\overline{I}^{2} V_{\mathrm{ref}}^{2}-(P^*)^{2}}
		-
		H\overline{I}^{2}
		\right)
	}{
		\overline{I}^{2} R_g
	}.
	\label{eq:cond_Vgb_minus}
\end{equation}

Unlike \(C^+\), the LBV boundary slope \(C^-\) can be positive, allowing stable operation of the converter once the limiter is activated. Both \(C^+\) and \(C^-\) depend only on the equivalent grid impedance, current limit, voltage reference, and active-power setpoint, and can therefore be evaluated without the dynamic operating states.
\subsection{Collapse Voltage}
The collapse voltage \(V_{g,\mathrm{col}}\) is the grid voltage at which the admissible operating equilibrium is lost. The lower and upper collapse voltages, \(V_{g,\mathrm{col}}^{-}\) and \(V_{g,\mathrm{col}}^{+}\), are abbreviated as the LCV and UCV, respectively. Since \(C^+<0\), the UCV coincides with the UBV:
\begin{equation}
	V_{g,\mathrm{col}}^{+}=V_{g,b}^{+}.
\end{equation}
At the LBV, \(C^-\leq0\) likewise gives
\begin{equation}
	V_{g,\mathrm{col}}^{-}=V_{g,b}^{-}.
\end{equation}
When \(C^->0\), a satSEP persists below the LBV, and the LCV is instead determined by the maximum active power that can be transferred in the CLM. Because the current magnitude is fixed at \(\overline{I}\), while its phase angle \(\phi\) is determined by the circuit equations, the saturated power can be written as
\begin{equation}
	\begin{aligned}
	&	P_{\mathrm{CLM}}(\delta,V_g)
	=
	P_{\mathrm{CLM}}\big(\phi(\delta,V_g),\delta,V_g\big)
	=
	\frac{R_g \overline{I}^2}{A}\\
	&+
	\frac{V_g \overline{I}}{A}
	\left[
	\left(1-B_cX_g\right)\cos\left(\phi+\delta\right)
	-
	B_cR_g\sin\left(\phi+\delta\right)
	\right].
	\label{eq:u}
\end{aligned}
\end{equation}
Here, \(\phi\) denotes the phase angle of the saturated inverter current phasor. It should be noted that \(\phi\) is not an independent state variable. For a given operating point, it is determined by the algebraic constraints of the equivalent circuit and can therefore be expressed as a function of \(\delta\) and \(V_g\). 

The existence of a steady-state solution in the CLM requires that the demanded active power \(P^*\) does not exceed the maximum active power that can be delivered under the current limit. 
From \eqref{eq:u}, the phase-dependent term satisfies
\begin{equation}
	\begin{aligned}
		&\max_{\phi+\delta}
		\left[
		\left(1-B_cX_g\right)\cos\left(\phi+\delta\right)
		-
		B_cR_g\sin\left(\phi+\delta\right)
		\right]  \\
		&=
		\sqrt{
			\left(1-B_cX_g\right)^2
			+
			\left(B_cR_g\right)^2
		}
		=
		\sqrt{A}.
	\end{aligned}
\end{equation}

Therefore, the maximum active power in CLM is
\begin{equation}
	P_{\mathrm{CLM,max}}(V_g)
	=
	\frac{R_g \overline{I}^2}{A}
	+
	\frac{V_g \overline{I}}{\sqrt{A}} .
	\label{eq:Pon_max}
\end{equation}
The LCV is reached when the maximum deliverable active power equals the setpoint:
\begin{equation}
	P_{\mathrm{CLM,max}}(V_g)
	=
	P^* .
\end{equation}
Substituting \eqref{eq:Pon_max} and solving for the LCV yields
\begin{equation}
	V_{g,\mathrm{col}}^{-}
	=
	\frac{
		A P^*
		-
		R_g \overline{I}^2
	}{
		\sqrt{A} \overline{I}
	},
	\qquad C^->0 .
	\label{eq:Vg_col_minus}
\end{equation}
The LCV can therefore be written as
\begin{equation}
	V_{g,\mathrm{col}}^{-}
	=
	\begin{cases}
		V_{g,b}^{-},
		&
		C^-\le 0,
		\\[2mm]
		\dfrac{
			A P^*
			-
			R_g \overline{I}^2
		}{
			\sqrt{A} \overline{I}
		},
		&
		C^->0.
	\end{cases}
	\label{eq:Vg_col_piecewise}
\end{equation}
Together, \eqref{eq:Vgb}, \eqref{eq:cond_Vgb_plus}, \eqref{eq:cond_Vgb_minus}, and \eqref{eq:Vg_col_piecewise} provide the key results used in the following analysis: when current limiting begins, whether a satSEP forms at the boundary, and when the operating equilibrium is lost.
\section{Application and Validation}\label{sec:application}
This section applies the analytical conditions to numerical examples and explains the predicted inverter behaviour using power-angle curves. The predictions are then validated through dynamic-model continuation and time-domain simulations in single-inverter and modified WSCC 9-bus systems.

\subsection{Numerical Evaluation and Power-Angle Analysis}

\begin{figure}[t]
    \centering
    \includegraphics[width=1\linewidth]{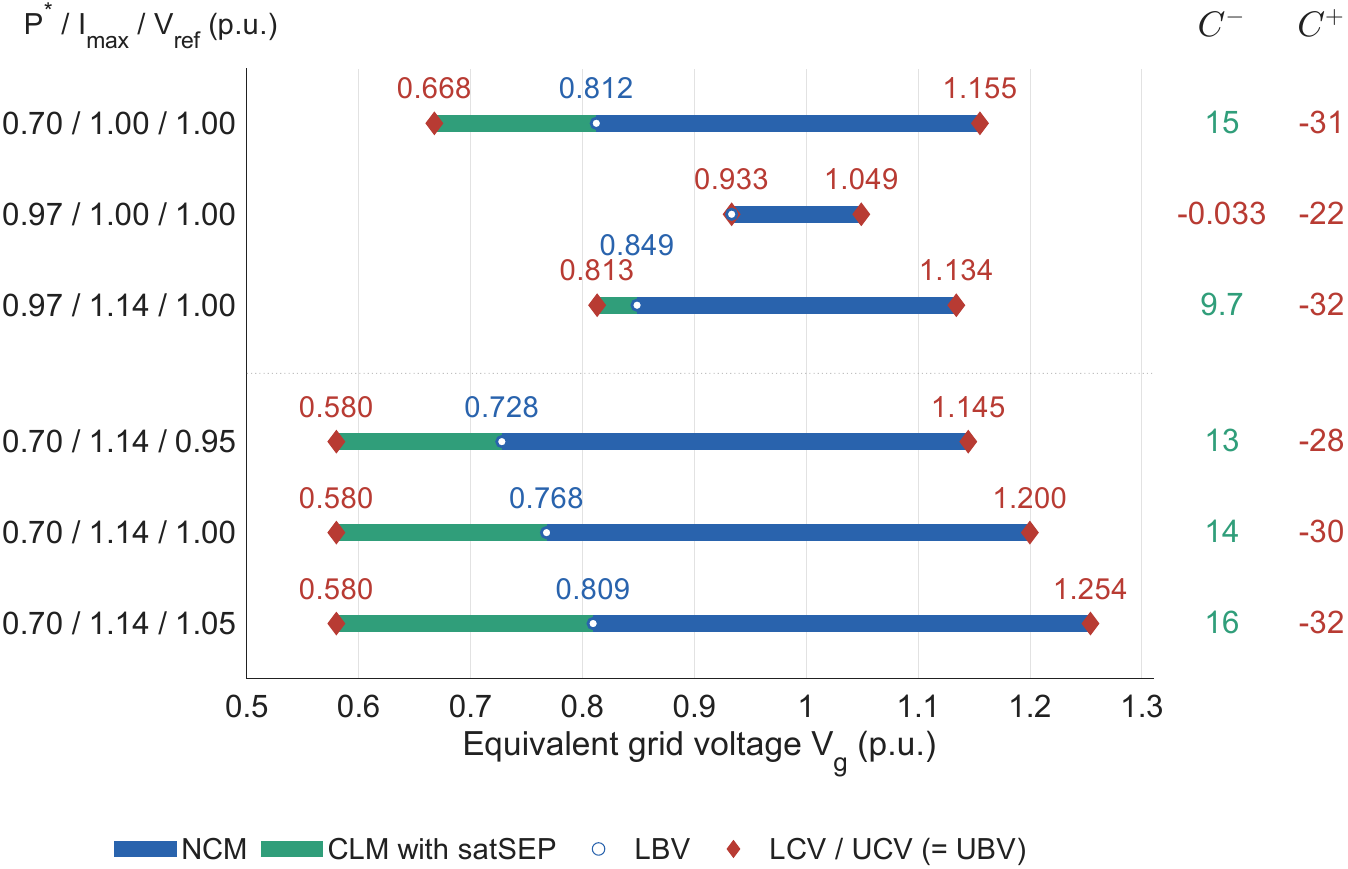}
    \caption{Boundary slopes and voltage thresholds for different active-power settings, current limits, and voltage references.}
    \label{tab:boundary_condition}
\end{figure}

The analytical conditions are evaluated for the SIIB system with \(R_g=0.021\), \(L_g=0.24\), and \(B_c=0.066\)~p.u. Two current limits, \(\overline{I}=1.0\) and \(1.14\)~p.u., are considered. The higher current limit is adopted from \cite{judge_power_nodate} based on the maximum allowable thermal loading of the converter. Two active-power setpoints are considered, and the higher setpoint of \(P^*=0.97\) is selected to illustrate a negative lower-boundary slope, \(C^-<0\), when \(\overline{I}=1.0\)~p.u. and \(V_{\mathrm{ref}}=1.0\)~p.u., demonstrating that equilibrium loss can occur immediately when CCL activation. Fig.~\ref{tab:boundary_condition} summarises the boundary slopes and voltage thresholds for six combinations of \(P^*\), \(\overline{I}\), and \(V_{\mathrm{ref}}\). The boundary slopes indicate whether the inverter can continue operating stably after the current limiter is activated.

In Fig.~\ref{tab:boundary_condition}, increasing the current limit lowers the LBV and raises the UBV, expanding the grid-voltage range for NCM operation. In all six cases, \(C^+<0\), so the UCV coincides with the UBV. At the lower boundary, \(C^->0\) for both current limits when \(P^*=0.70\) and \(V_{\mathrm{ref}}=1.0\) p.u.,indicating that the inverter can continue operating after the current limiter is activated. For \(P^*=0.97\), increasing \(\overline{I}\) from \(1.0\) to \(1.14\)~p.u. changes \(C^-\) from negative to positive. Thus, increasing the current limit and reducing power setpoints can extend the voltage operating range.

In practical applications, the effective voltage reference may vary through \(Q\)--\(V\) droop control or a virtual impedance. The three cases with \(P^*=0.70\) and \(\overline{I}=1.14\) show that both the LBV and UBV shift downward when \(V_{\mathrm{ref}}\) is reduced and upward when it is increased. These results indicate how voltage-reference adjustments can delay CCL activation: a lower reference reduces the LBV during voltage sags, while a higher reference raises the UBV during voltage rises.

The two cases with \(\overline{I}=1.0\)~p.u. are examined further using power-angle curves to explain why stable operation can continue after current-limiter activation for \(P^*=0.70\), but is lost at activation for \(P^*=0.97\). To obtain each \(P\)-\(\delta\) curve, for each specified value of \(V_g\), the angle \(\delta\) is varied from \(0^\circ\) to \(180^\circ\), and the corresponding active power is calculated from the piecewise expression in \eqref{eq:P_with_cap}.

\subsubsection{Decreasing Grid Voltage}

\begin{figure*}[t]
	\centering
	\begingroup
	
	\setlength{\tabcolsep}{0pt}
	\captionsetup[subfigure]{skip=0pt}
	
	\def\pdeltapanel#1{%
		\includegraphics[width=\linewidth]{#1}%
	}
	
	\begin{tabular}{
		@{}
		>{\centering\arraybackslash}m{0.492\textwidth}
		@{\hspace{0.016\textwidth}}
		>{\centering\arraybackslash}m{0.492\textwidth}
		@{}
	}
		\textbf{Decreasing grid voltage}
		&
		\textbf{Increasing grid voltage}
		\\[0.15ex]
		
		{\setcounter{subfigure}{0}%
		\subcaptionbox{\(P^*=0.70\)\label{fig:pdeltavgdecreasingpset07withvgcol}}
		[0.492\textwidth]{%
			\pdeltapanel{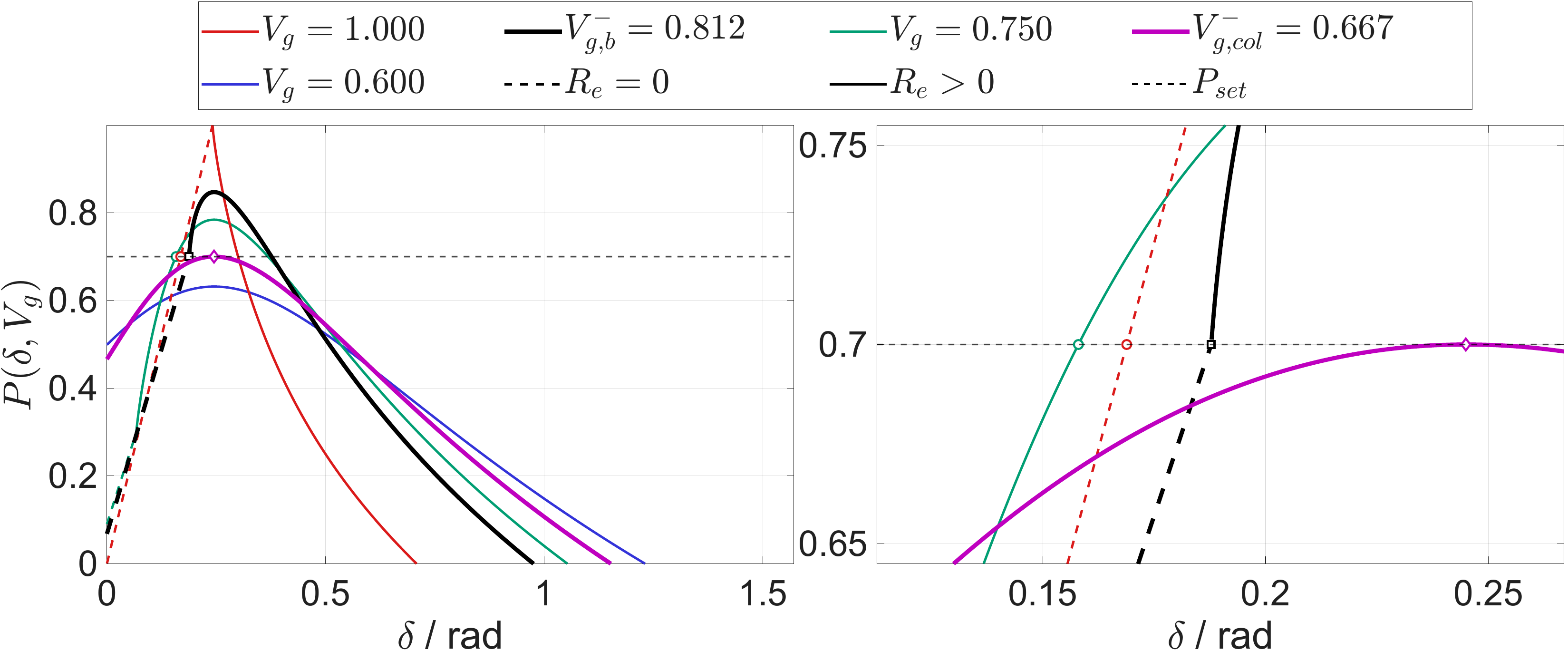}
		}}
		&
		{\setcounter{subfigure}{2}%
		\subcaptionbox{\(P^*=0.70\)\label{fig:Pdelta_Vg_rising_combined}}
		[0.492\textwidth]{%
			\pdeltapanel{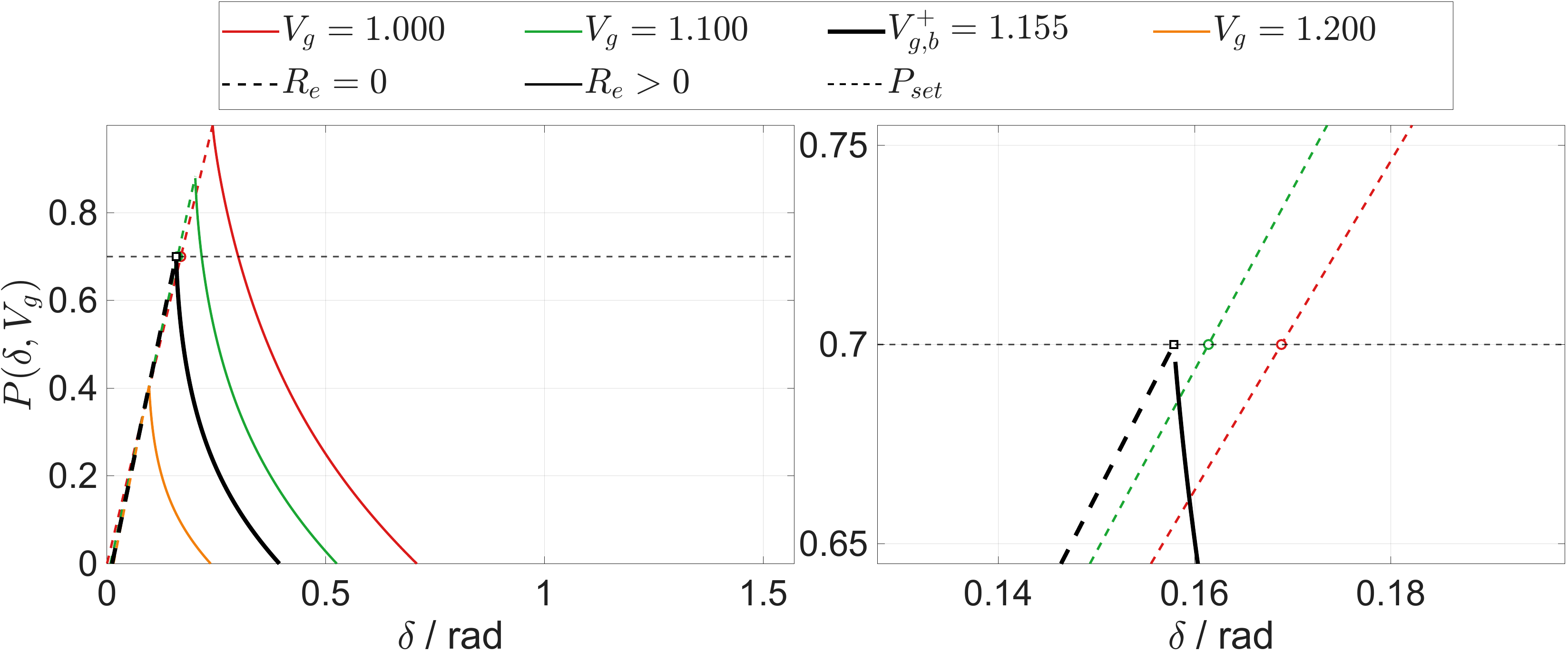}
		}}
		\\[-0.15ex]
		
		{\setcounter{subfigure}{1}%
		\subcaptionbox{\(P^*=0.97\)\label{fig:pdeltavgdecreasingpset097}}
		[0.492\textwidth]{%
			\pdeltapanel{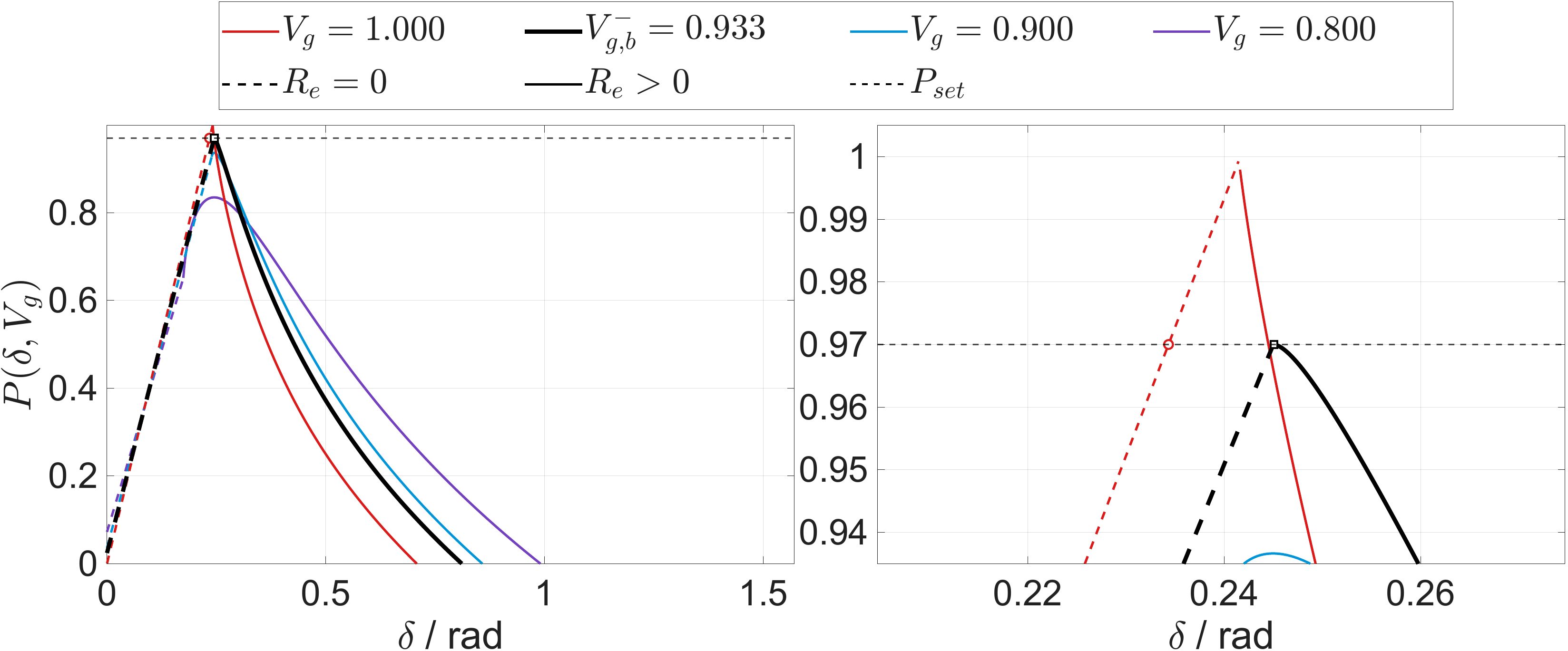}
		}}
		&
		{\setcounter{subfigure}{3}%
		\subcaptionbox{\(P^*=0.97\)\label{fig:Pdelta_Vg_rising_Pset097_combined}}
		[0.492\textwidth]{%
			\pdeltapanel{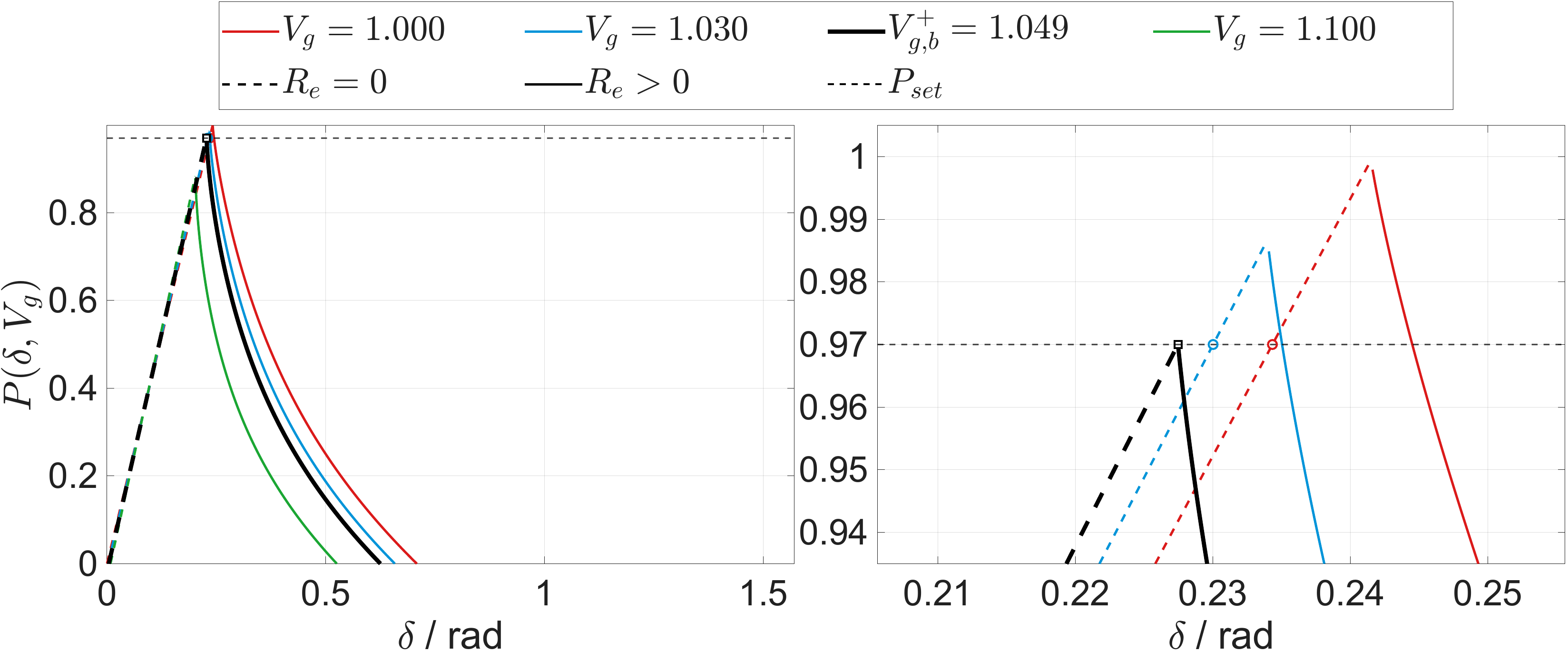}
		}}
	\end{tabular}
	
	\endgroup
	
	\caption{\(P\)--\(\delta\) curves under decreasing and increasing grid voltage for two active-power setpoints. Panels (a) and (b) correspond to decreasing grid voltage with \(P^*=0.70\) and \(P^*=0.97\), respectively, whereas panels (c) and (d) correspond to increasing grid voltage with \(P^*=0.70\) and \(P^*=0.97\), respectively. In each panel, the right-hand plot shows a zoomed view of the corresponding region in the left-hand plot.}
	\label{fig:Pdelta_Vg_combined_all}
\end{figure*}

Figs.~\ref{fig:pdeltavgdecreasingpset07withvgcol} and~\ref{fig:pdeltavgdecreasingpset097} show the evolution of the \(P\)-\(\delta\) characteristic as the grid voltage is reduced. The reduction in \(V_g\) lowers the maximum transferable active power of both the NCM and CLM portions of the curve.

For \(P^*=0.70\), the operating intersection with \(P=P^*\) initially lies on the NCM part of the curve. As \(V_g\) decreases, this intersection reaches the switching boundary at the LBV. The positive saturated boundary slope allows the intersection to continue into the CLM, so a satSEP persists below the LBV.

As \(V_g\) is reduced further, the maximum value of the saturated \(P\)-\(\delta\) curve continues to decrease. At the LCV, this maximum becomes equal to \(P^*\), corresponding to the saddle node bifurcation point predicted by \eqref{eq:Vg_col_minus}. Below the LCV, the saturated curve no longer intersects the demand line, and the satSEP disappears.

For \(P^*=0.97\), the same curve family used for \(P^*=0.70\) is intersected by a higher demand line as shown in Fig.~\ref{fig:pdeltavgdecreasingpset097}. In this case, the saturated boundary slope is negative,  so no satSEP beyond LBV. Consequently, the LCV coincides with the LBV, and the operating equilibrium is lost upon CCL activation, corresponding to a non-smooth fold.

\begin{figure*}[t]
    \centering
    \begin{minipage}[t]{0.47\linewidth}
        \centering
        \includegraphics[width=\linewidth]{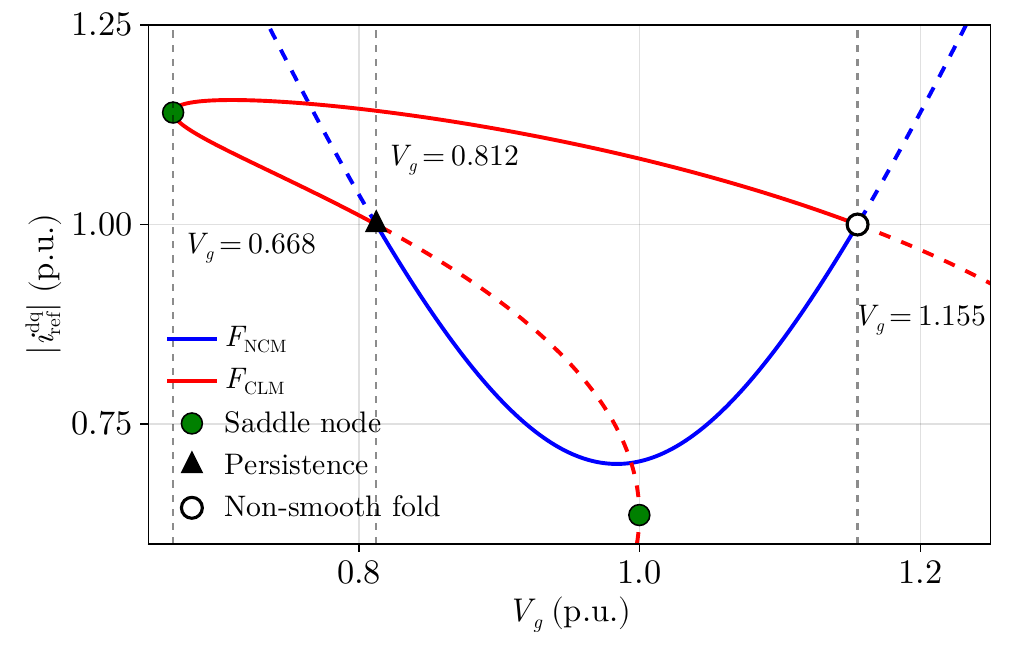}
        \caption{Bifurcation diagram for Case~1 at \(P^*=0.70\).}
        \label{fig:Imag_BEB_0.7}
    \end{minipage}
    \hfill
    \begin{minipage}[t]{0.47\linewidth}
        \centering
        \includegraphics[width=\linewidth]{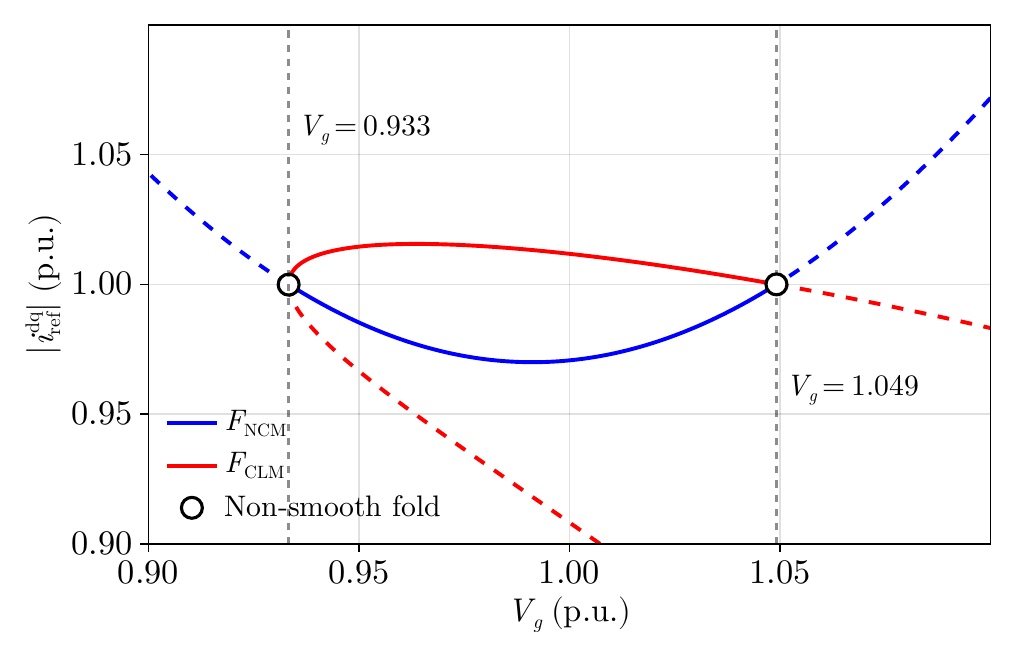}
        \caption{Bifurcation diagram for Case~1 at \(P^*=0.97\).}
        \label{fig:Imag_BEB_0.97}
    \end{minipage}
\end{figure*}
\subsubsection{Increasing Grid Voltage}

Figs.~\ref{fig:Pdelta_Vg_rising_combined} and~\ref{fig:Pdelta_Vg_rising_Pset097_combined} show the corresponding curves for increasing grid voltage. For both active-power settings, Fig.~\ref{tab:boundary_condition} gives \(C^+<0\). This agrees with \eqref{eq:cond_Vgb_plus}, which shows that the UBV boundary slope is always negative under the assumed parameter signs. Hence, no satSEP exists beyond the UBV.

As \(V_g\) increases, the operating point remains on the NCM part of the \(P\)-\(\delta\) curve until the UBV. At this boundary, the CCL is activated, but the saturated curve does not intersect \(P=P^*\) at an admissible equilibrium. Therefore, \(\mathrm{UCV}=\mathrm{UBV}\): when increasing grid voltage activates the CCL at the UBV, the operating equilibrium is lost immediately.
\subsection{Validation Using Bifurcation Analysis of the Dynamic Model}\label{sec:result}

\begin{table}[t]
	\centering
	\caption{GFM parameters in their respective per-unit bases.}
	\label{tab:gfm_parameters}
	\begin{tabular}{lll}
		\toprule
		Symbol & Description & Value \\
		\midrule
		$m_p$ & Frequency droop gain & $0.02$ \\
		$\omega_c$ & Cut-off frequency of the filter & $31.4$ \\
		$\omega_g$ & Nominal angular frequency in p.u. & $1$ \\
		$\omega_b$ & Base angular frequency & $60 \times 2\pi$ \\
        \midrule
        $V_{\rm ref}$ & Reference voltage & 1 \\
		$\overline{I}$ & Continuous maximum current magnitude & 1\\
		\midrule
		$L_f$ & Filter inductance & $0.15$ \\
		$C_f$ & Filter capacitance & $0.066$ \\
		$R_f$ & Filter resistance & $0$ \\
		\midrule
		$K_p^v$ & Voltage controller proportional gain & $0.52$ \\
		$K_i^v$ & Voltage controller integral gain & $1.16$ \\
		\midrule
		$K_p^i$ & Current controller proportional gain & $0.73$ \\
		$K_i^i$ & Current controller integral gain & $1.19$ \\
		\bottomrule
	\end{tabular}
\end{table}
The preceding power-angle analysis is based on the equivalent-circuit model. This subsection validates the analytical predictions through equilibrium continuation of the full dynamic models in three test systems. The full dynamic models are implemented in PowerDynamics.jl \cite{plietzsch_powerdynamicsjlexperimentally_2022}, while the bifurcation analyses are performed using BifurcationKit.jl \cite{veltz_bifurcationkitjl_2020} with the pseudo-arclength continuation method. All GFM converters use the control structure shown in Fig.~\ref{fig:gfm_control_structure}, with the controller parameters in Table~\ref{tab:gfm_parameters} adopted from \cite{rokrok_classification_2020}. The SGs are modelled as balanced eighth-order machines equipped with ST1A excitation systems \cite{sauer_power_nodate}. 

Case~1 is the SIIB system. Its grid parameters are \(R_g=0.021\) and \(L_g=0.24\)~p.u. adopted from \cite{fan_equivalent_2022}, and the infinite-bus voltage magnitude is used as the primary continuation parameter. Case~2 is a modified Western System Coordinating Council (WSCC) 9-bus system with an infinite bus at bus~1, a GFM inverter at bus~2, and an SG at bus~3. The infinite-bus voltage magnitude is used as the primary continuation parameter. Case~3 is a modified WSCC 9-bus system with SGs at buses~1 and~3 and a GFM inverter at bus~2. The nominal active-power demand at bus~9 is used as the primary continuation parameter. 

In Cases~2 and~3, the Thevenin parameters \(R_g\) and \(L_g\) are calculated using small perturbations at the inverter terminal at each equilibrium point obtained by continuation. Their values depend on the system operating condition. The corresponding grid voltage is reconstructed from the inverter terminal voltage and output current as
\begin{equation}
	V_g = |V_f - I_g (R_g + jX_g)|.\label{eq:reconstructed_grid_voltage}
\end{equation}

\subsubsection{Single-Parameter Validation at Two Active-Power Setpoints}\label{sec:single_para}

To validate the analytical predictions, the equilibria of the full dynamic model in Case~1 are analyzed using bifurcation analysis, following the method described in Section~\ref{sec:reduced_saturated_formulation}. The continuation examines whether the predicted equilibrium behaviour occurs at the calculated voltage thresholds. The resulting bifurcation diagrams are shown in Figs.~\ref{fig:Imag_BEB_0.7} and~\ref{fig:Imag_BEB_0.97}, where solid and dashed branches denote admissible and virtual equilibria, respectively.

For both setpoints, the CCL is activated at the analytically predicted LBV and UBV in Fig.~\ref{tab:boundary_condition}. For \(P^*=0.70\), the admissible equilibrium continues into the CLM at $V_g = 0.812$ and is lost at a later saddle node whose location agrees with the predicted LCV. This behaviour corresponds to persistence, meaning that the inverter can continue operating stably after the current limiter is activated. For \(P^*=0.97\), no admissible equilibrium remains beyond the LBV at $V_g = 0.933$, so the inverter loses stability upon limiter activation. This corresponds to a non-smooth fold. At the upper boundary, both setpoints lead to immediate equilibrium loss and converter instability when the current limiter is activated, also corresponding to a non-smooth fold.

\begin{figure*}[t]
    \centering
    \begingroup

    \setlength{\tabcolsep}{0pt}
    \captionsetup[subfigure]{skip=0pt}

    \newlength{\caselabelwidth}
    \newlength{\casegap}
    \newlength{\panelgap}
    \newlength{\panelwidth}

    \setlength{\caselabelwidth}{0.025\textwidth}
    \setlength{\casegap}{0.010\textwidth}

    \setlength{\panelgap}{0.040\textwidth}
    \setlength{\panelwidth}{0.440\textwidth}

    \newcommand{\voltagepanel}[1]{%
        \includegraphics[
            width=\linewidth
        ]{#1}%
    }

    \begin{tabular}{
        @{}
        >{\centering\arraybackslash}m{\caselabelwidth}
        @{\hspace{\casegap}}
        >{\centering\arraybackslash}m{\panelwidth}
        @{\hspace{\panelgap}}
        >{\centering\arraybackslash}m{\panelwidth}
        @{}
    }

        &
        \textbf{Boundary voltages}
        &
        \textbf{Collapse voltages}
        \\[0ex]

        \rotatebox[origin=c]{90}{\textbf{Case~1}}
        &
        {%
            \setcounter{subfigure}{0}%
            \subcaptionbox{%
                \label{fig:case1_boundary_voltage}%
            }[\linewidth]{%
                \voltagepanel{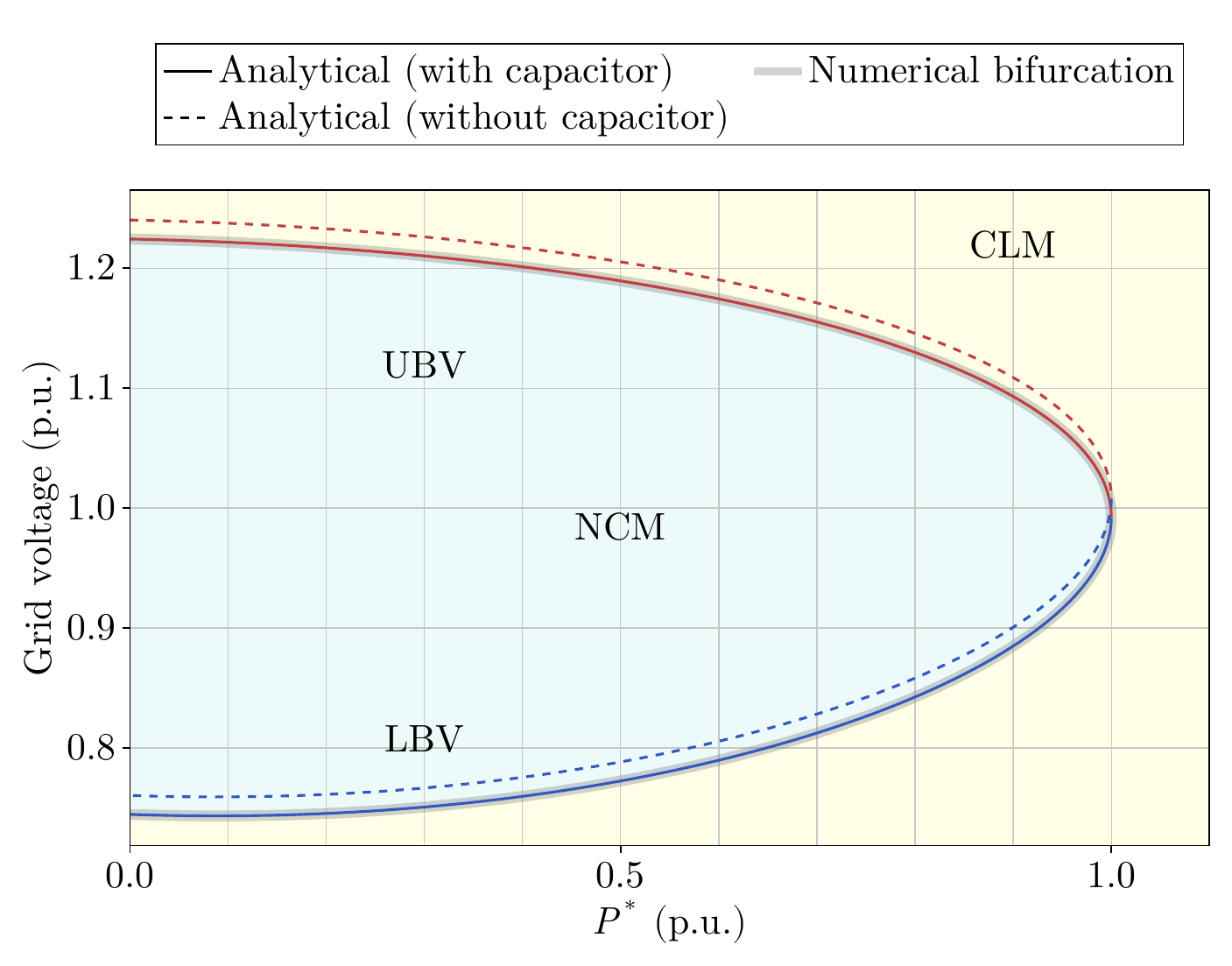}%
            }%
        }
        &
        {%
            \setcounter{subfigure}{3}%
            \subcaptionbox{%
                \label{fig:case1_collapse_voltage}%
            }[\linewidth]{%
                \voltagepanel{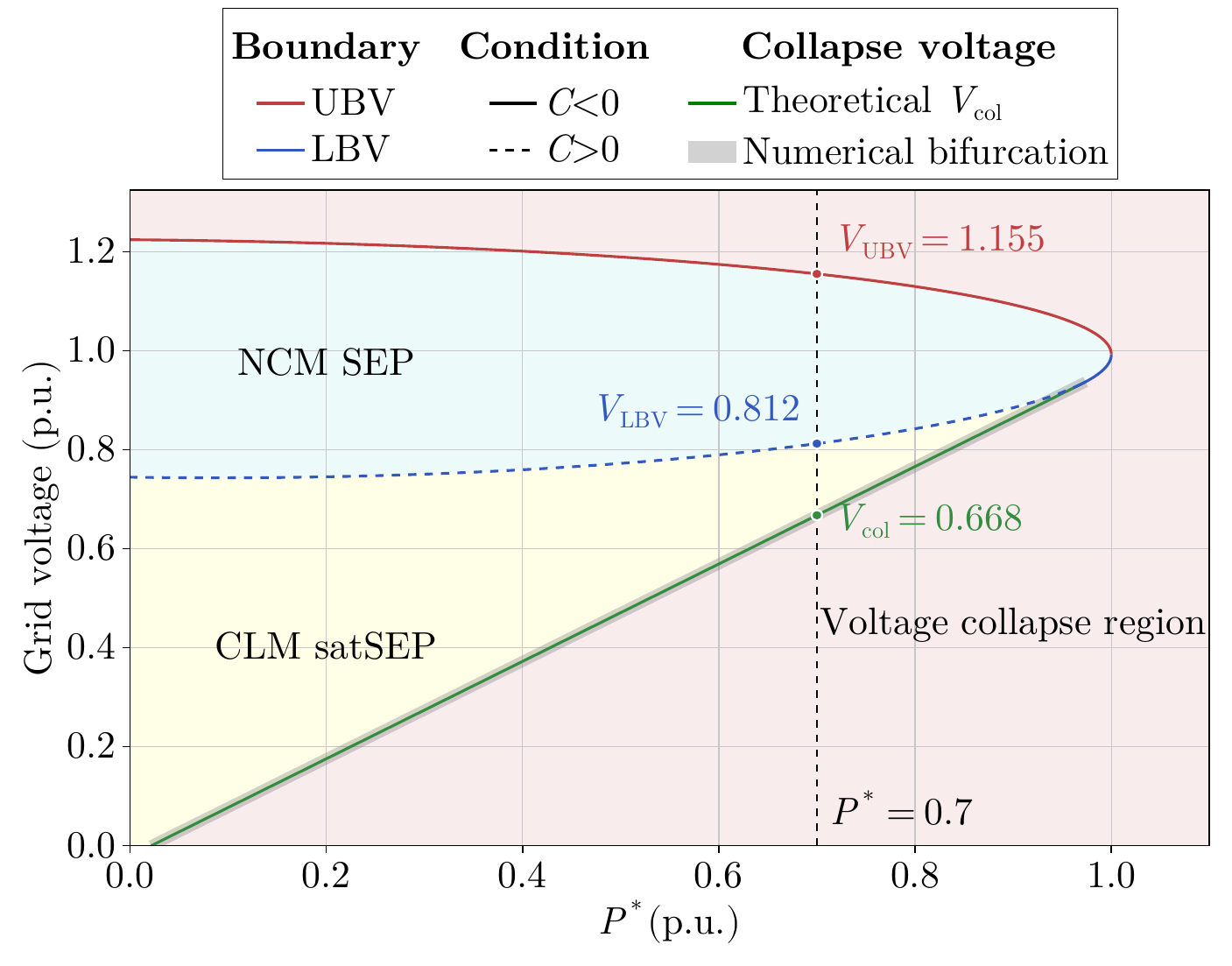}%
            }%
        }
        \\[-0.5ex]

        \rotatebox[origin=c]{90}{\textbf{Case~2}}
        &
        {%
            \setcounter{subfigure}{1}%
            \subcaptionbox{%
                \label{fig:case2_boundary_voltage}%
            }[\linewidth]{%
                \voltagepanel{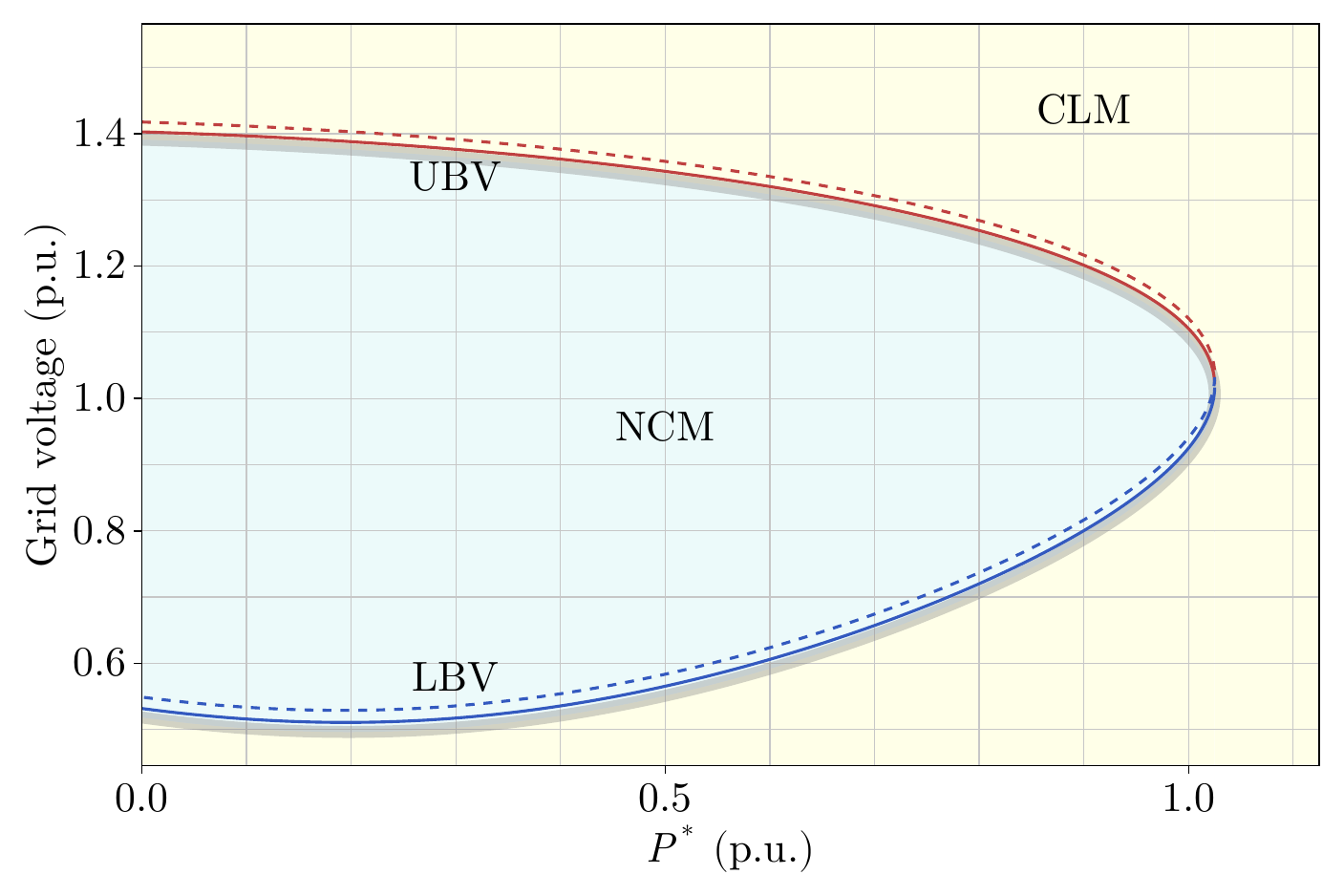}%
            }%
        }
        &
        {%
            \setcounter{subfigure}{4}%
            \subcaptionbox{%
                \label{fig:case2_collapse_voltage}%
            }[\linewidth]{%
                \voltagepanel{images/Case9_inf_collapse_voltage.pdf}%
            }%
        }
        \\[-0.5ex]

        \rotatebox[origin=c]{90}{\textbf{Case~3}}
        &
        {%
            \setcounter{subfigure}{2}%
            \subcaptionbox{%
                \label{fig:case3_boundary_voltage}%
            }[\linewidth]{%
                \voltagepanel{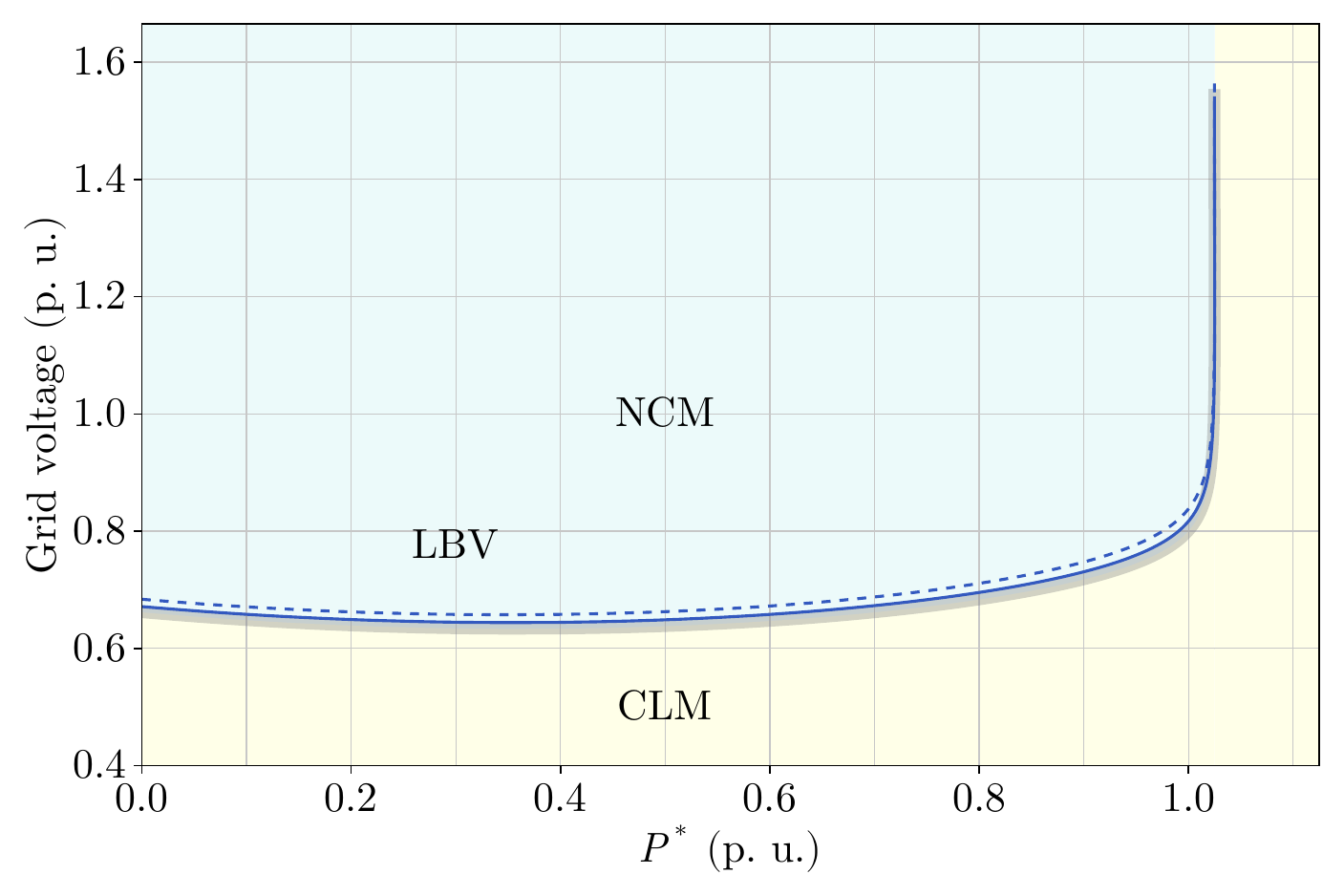}%
            }%
        }
        &
        {%
            \setcounter{subfigure}{5}%
            \subcaptionbox{%
                \label{fig:case3_collapse_voltage}%
            }[\linewidth]{%
                \voltagepanel{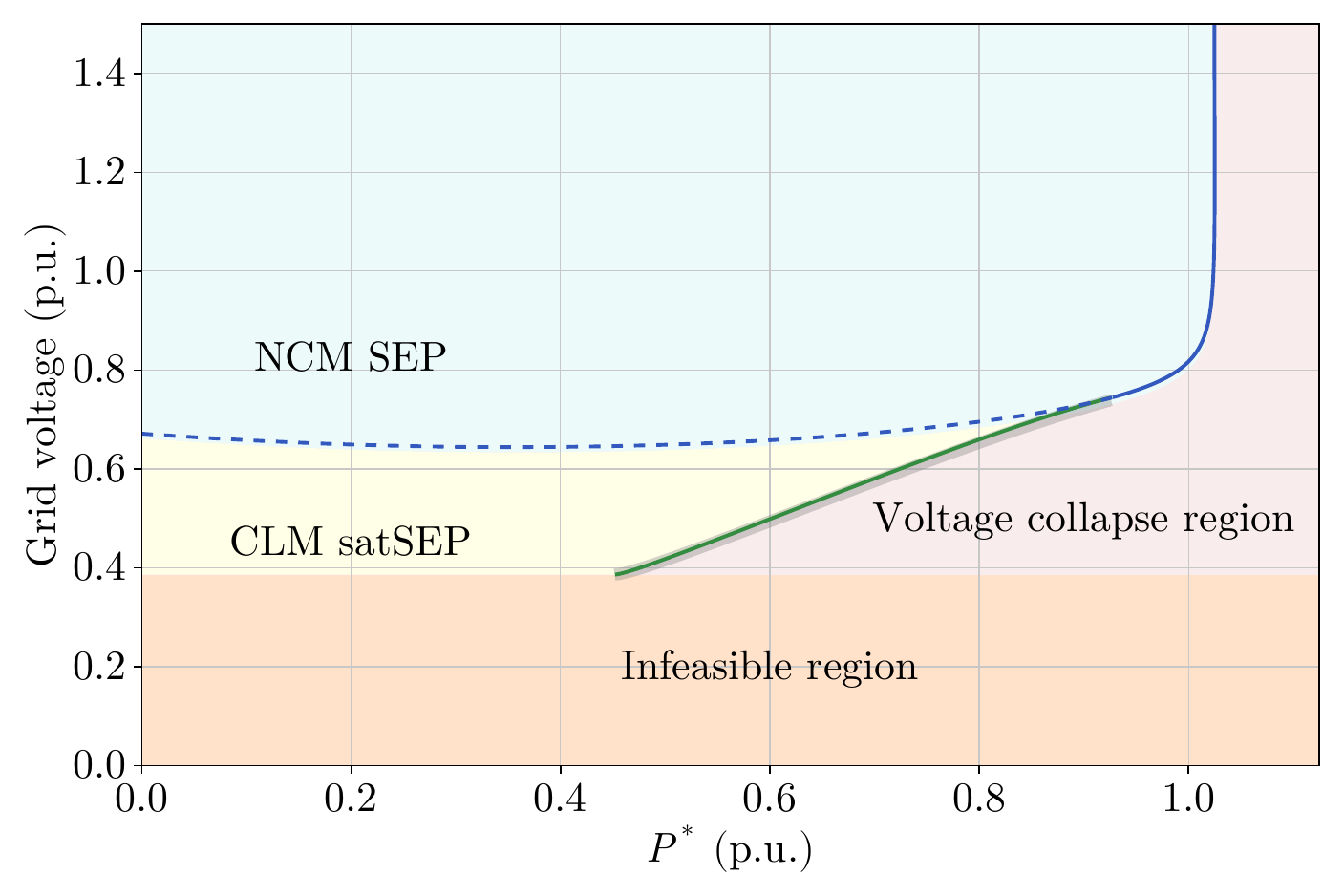}%
            }%
        }

    \end{tabular}

    \endgroup

    \caption{Two-parameter boundary- and collapse-voltage results for the three studied cases.}
    \label{fig:two_parameter_voltage_results}
\end{figure*}

\begin{figure*}[t]
	\centering
	\begingroup
	
	\setlength{\tabcolsep}{0pt}
	\captionsetup[subfigure]{skip=0pt}
	
	\def\timepanel#1{%
		\includegraphics[width=\linewidth]{#1}%
	}
	
	\begin{tabular}{
		@{}
		>{\centering\arraybackslash}m{0.492\textwidth}
		@{\hspace{0.016\textwidth}}
		>{\centering\arraybackslash}m{0.492\textwidth}
		@{}
	}
		\textbf{Increasing infinite-bus voltage}
		&
		\textbf{Decreasing infinite-bus voltage}
		\\[-0.8ex]
		
		{\setcounter{subfigure}{0}%
		\subcaptionbox{\label{fig:increase_iunorm}}
		[0.492\textwidth]{%
			\timepanel{
				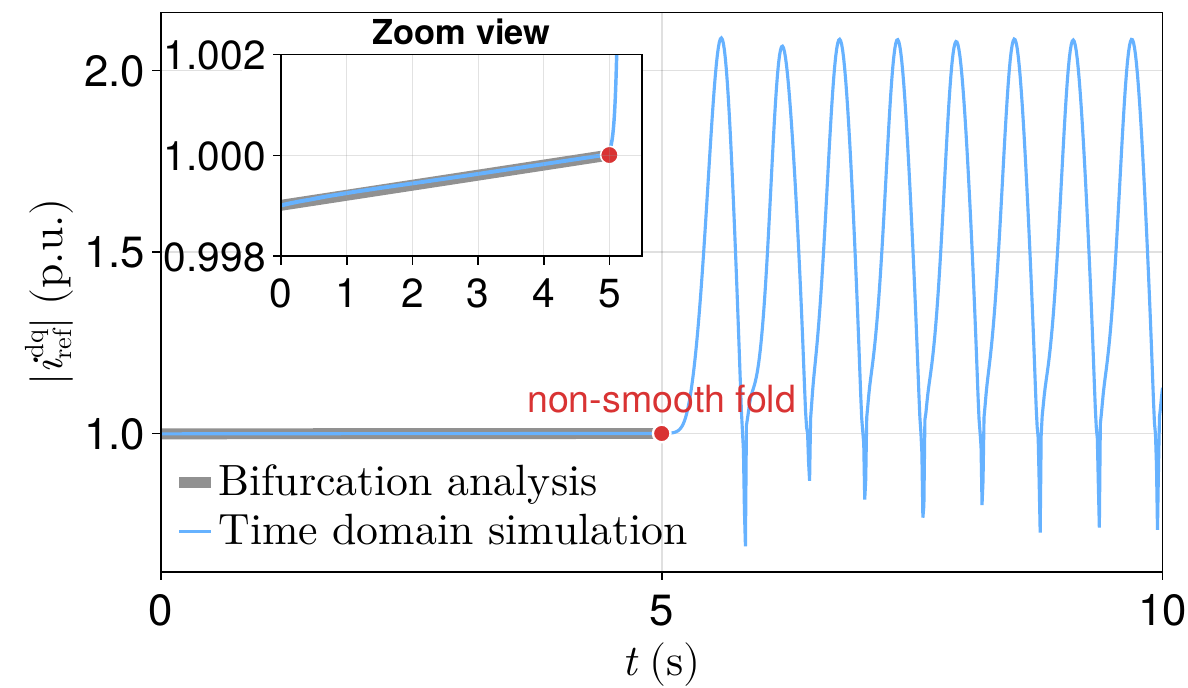
			}
		}}
		&
		{\setcounter{subfigure}{2}%
		\subcaptionbox{\label{fig:decrease_iunorm}}
		[0.492\textwidth]{%
			\timepanel{
				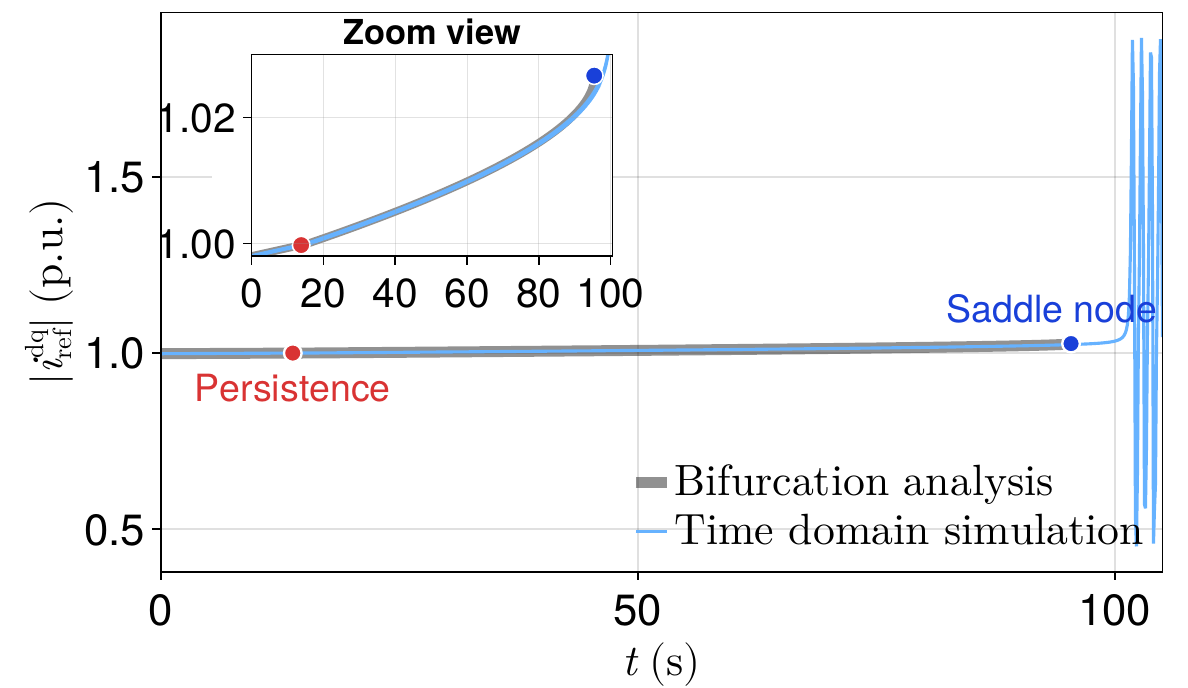
			}
		}}
		\\[-0.0ex]
		
		{\setcounter{subfigure}{1}%
		\subcaptionbox{\label{fig:increase_v2}}
		[0.492\textwidth]{%
			\timepanel{
				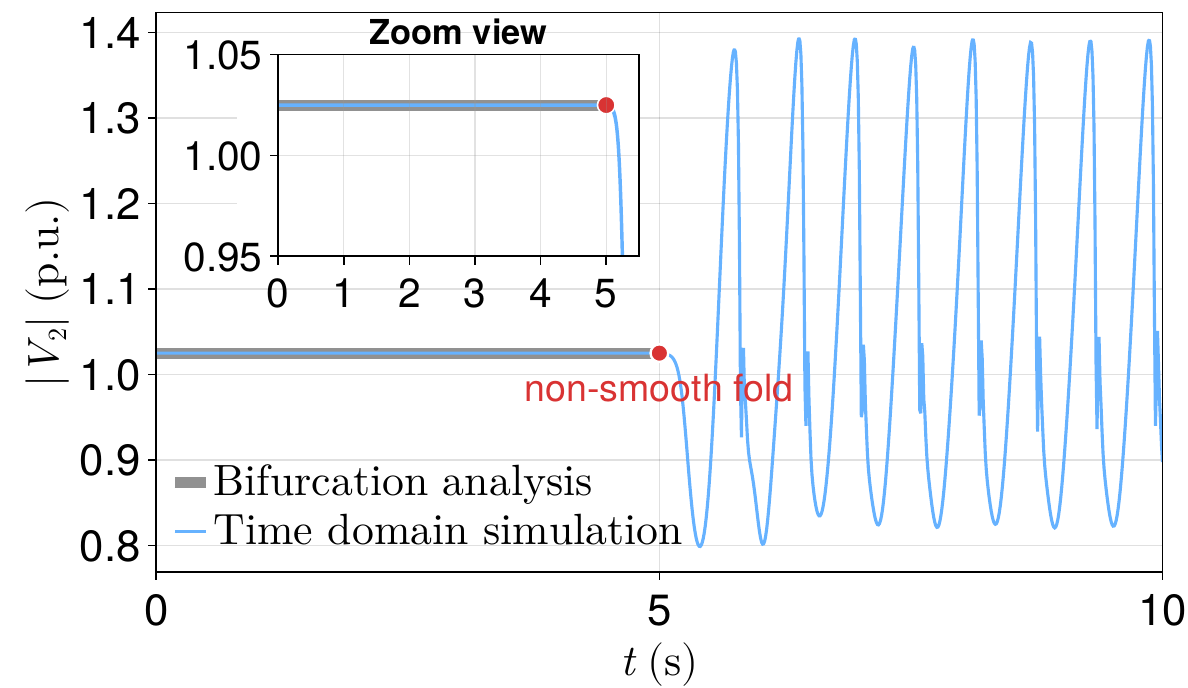
			}
		}}
		&
		{\setcounter{subfigure}{3}%
		\subcaptionbox{\label{fig:decrease_v2}}
		[0.492\textwidth]{%
			\timepanel{
				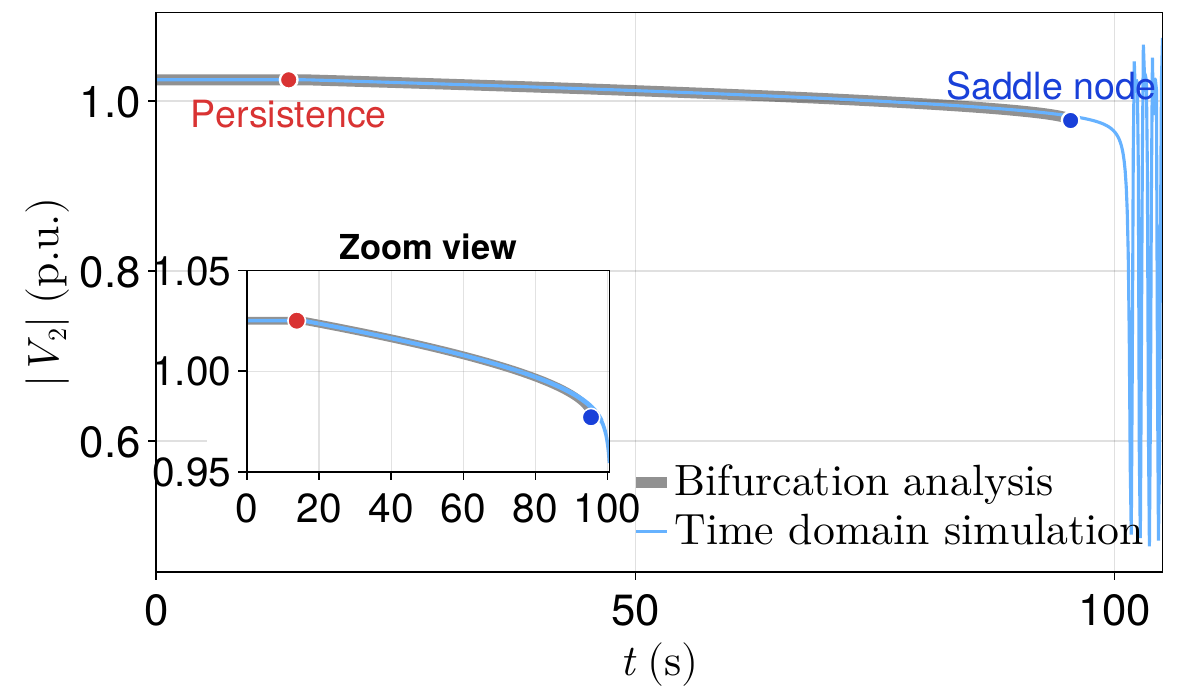
			}
		}}
	\end{tabular}
	
	\endgroup
	
\caption{Time-domain responses of Case~2: unconstrained current-reference and inverter-bus voltage magnitudes under (a)--(b) increasing infinite-bus voltage, causing instability at limiter activation, and (c)--(d) decreasing infinite-bus voltage, allowing stable current-limited operation until a subsequent saddle node.}	\label{fig:time_domain_inf_voltage_variation}
\end{figure*}
\subsubsection{Two-Parameter Validation of Boundary Voltages}
The analytical expression in~\eqref{eq:Vgb} predicts how the LBV and UBV vary with \(P^*\). To validate these predictions over a range of active-power setpoints, the boundary voltage points of the full dynamic model are continued with \(P^*\) as the second parameter. The results for the three cases are shown in Fig.~\ref{fig:two_parameter_voltage_results}~(a)--(c). The region between the LBV and UBV corresponds to the NCM, whereas the regions outside these boundaries correspond to the CLM. As \(P^*\) increases, the two boundaries generally move closer together, narrowing the grid-voltage range over which the inverter remains in the NCM.

For comparison, \eqref{eq:Vgb} is evaluated both with the actual filter-capacitor susceptance \(B_c\) and with \(B_c=0\), the latter corresponding to the equivalent-circuit approximation that neglects the filter capacitor \cite{fan_equivalent_2022}. The curves obtained using the actual \(B_c\) closely follow the dynamic-model continuation results, whereas setting \(B_c=0\) shifts the calculated boundary voltages upward. In case~1, the root-mean-square error of the boundary-voltage predictions is \(5.3\times10^{-15}\)~p.u. when the capacitor is included, compared with \(1.6\times10^{-2}\)~p.u. when it is neglected. The analytical predictions therefore match the dynamic-model bifurcation results to numerical precision, allowing the onset of current limiting to be calculated directly without sacrificing accuracy in this case.
\subsubsection{Two-Parameter Validation of Collapse Voltage}

Starting from the saddle node identified by single-parameter continuation in Section~\ref{sec:single_para}, two-parameter continuation with \(P^*\) traces the variation of the LCV with the active-power setpoint.

Fig.~\ref{fig:two_parameter_voltage_results}~(d)--(f) compares the analytical collapse voltage with the continuation results of the dynamic models. Where the equilibrium is lost upon CCL activation, the collapse voltage coincides with the LBV or UBV. When the equilibrium persists after CCL activation, the dashed LBV marks only the activation boundary, and equilibrium loss occurs at the later LCV shown by the green curve. 

For example, in Case~1 with \(P^*=0.70\), Fig.~\ref{fig:two_parameter_voltage_results}(d) predicts equilibrium loss when the grid voltage exceeds \(1.155\)~p.u. Under decreasing grid voltage, the inverter enters the CLM at \(0.812\)~p.u., and a satSEP persists until the voltage reaches \(0.668\)~p.u. Below this value, the satSEP disappears. These thresholds agree with Fig.~\ref{tab:boundary_condition}. For all three cases, the boundary and saddle node continuation results closely follow the corresponding analytical curves, confirming whether equilibrium loss occurs at CCL activation or at a subsequent saddle node.
\subsection{Time-Domain Verification of the Two BEB Scenarios}
\label{sec:time_domain_verification}

Static continuation identifies the equilibrium branches but does not show how the PWS system responds when a boundary is crossed. Time-domain simulations of~\eqref{eq:piecewise_limited_model} are therefore performed for Case~2 with \(P^*=0.70\). A ramp increasing infinite-bus voltage drives the system through the UBV, where \(C^+<0\) predicts a non-smooth fold. A decreasing infinite-bus voltage drives it through the LBV, where \(C^->0\) predicts persistence into the CLM and a later saddle node. In Fig.~\ref{fig:time_domain_inf_voltage_variation}, grey denotes the admissible equilibrium branch of the dynamic model mapped to time through the imposed voltage trajectory, and blue denotes the simulated response. The voltage magnitude at the GFM bus is denoted by \(|V_2|\).

Using the reduced CLM model, saddle-node continuation obtains the collapse voltages for 89 different active-power setpoints shown in Fig.~\ref{fig:two_parameter_voltage_results}(e) in a total computation time of 8.59~s. In comparison, the time-domain analysis shown in Fig.~\ref{fig:time_domain_inf_voltage_variation}(c)--(d) requires 1.57~s to determine the collapse voltage for one setpoint. The proposed continuation method is therefore approximately 16 times faster. This comparison demonstrates the improved computational efficiency of the proposed continuation method.
\subsubsection{Non-Smooth Fold Under Increasing Infinite-Bus Voltage}

In Fig.~\ref{fig:time_domain_inf_voltage_variation} (a)(b), the simulated magnitude of the unconstrained current reference and the bus voltage track the NCM equilibrium up to the upper BEB at \(t=5\)~s, where \(|i^{\rm dq}_{\rm ref}|=\overline{I}\). Although the signals remain continuous when the CCL is activated, the response immediately leaves the equilibrium branch because no admissible CLM equilibrium exists beyond this boundary. The ensuing large excursions in \(|i^{\rm dq}_{\rm ref}|\) and \(|V_2|\) confirm the predicted non-smooth fold and show that the UBV is also the collapse voltage: current-limiter activation leads directly to converter instability, so the UBV is also the collapse voltage.

\subsubsection{Persistence Under Decreasing Infinite-Bus Voltage}
In Fig.~\ref{fig:time_domain_inf_voltage_variation} (c)(d), the unconstrained-current magnitude first reaches \(\overline{I}\) at the LBV near \(t=13.8\)~s. Unlike the increasing-voltage case, the simulated response does not leave the equilibrium branch. It passes continuously from the NCM branch to the admissible CLM branch and continues to track the grey curve, confirming the predicted persistence. This means that the inverter continues operating stably with the current limiter activated.

As the infinite-bus voltage decreases further, the CLM equilibrium reaches the saddle node near \(t=95.4\)~s. The parameter variation is stopped at this point, and the infinite-bus voltage is held constant. Nevertheless, the simulated response gradually departs from the saddle node and begins to oscillate, with large oscillations appearing after approximately \(t=102\)~s.

\section{Conclusion}
\label{sec:conclusion}

This paper presents an analytical method for assessing equilibrium existence in grid-forming inverters with circular current limiting and frozen anti-windup. The method predicts when current limiting is activated, whether an operating equilibrium remains available after activation, and the voltage at which that equilibrium is lost. An equivalent-circuit formulation that includes the filter capacitor yields closed-form expressions for the activation boundaries and collapse voltage. A slope condition determines whether the equilibrium is lost when current limiting is activated or persists until a later saddle-node bifurcation. A numerical continuation method enables equilibrium branches to be traced across limiter activation. The predictions are supported by power-angle analysis, dynamic-model continuation in single-inverter and modified WSCC 9-bus systems, and time-domain simulations.

The results show that current-limiter activation does not necessarily imply equilibrium loss. The proposed method helps identify operating conditions under which an inverter can sustain current-limited operation. Its analytical expressions offer potential for real-time monitoring and early warning of an approaching collapse as grid conditions change. Such warnings could guide preventive actions, including active-power curtailment and voltage-reference adjustments, to maintain an adequate operating margin. Future work will investigate online implementation and assess how these method can be coordinated with existing inverter controls.
\balance
\printbibliography
\end{document}